\documentclass[prd,showpacs]{revtex4}

\usepackage{amsmath}
\usepackage{amsthm}
\usepackage{amssymb}
\usepackage{amscd}
\usepackage[british]{babel}
\usepackage{graphicx}
\usepackage{psfrag}
\usepackage{epsfig}
\usepackage{rotating}
\usepackage{times}

\begin{document}

\title{Topics in Quantum Field Theory in Curved Space}

\author{Jaume Haro$^{1,}$\footnote{E-mail: jaime.haro@upc.edu}} 

\affiliation{$^1$Departament de Matem\`atica Aplicada I, Universitat
Polit\`ecnica de Catalunya, Diagonal 647, 08028 Barcelona, Spain}

\pagestyle{myheadings}

\theoremstyle{plain}
\newtheorem{lemma}{Lemma}[section]
\newtheorem{theorem}{Theorem}[section]
\newtheorem{proposition}{Proposition}[section]
\newtheorem{corollary}{Corollary}[section]
\newtheorem{remark}{Remark}[section]
\newtheorem{definition}{Definition}[section]
\newtheorem{example}{Example}[section]

\newcommand{\boxend}{\flushright{$\Box$}}
\newenvironment{dem}[1]{\begin{trivlist} \item {\bf Proof #1\/:}}
            {\boxend \end{trivlist}}

\newcommand{\N}{{\mathbb N}}               
\newcommand{\Z}{{\mathbb Z}}               
\newcommand{\Q}{{\mathbb Q}}               
\newcommand{\R}{{\mathbb R}}               
\newcommand{\C}{{\mathbb C}}               
\renewcommand{\S}{{\mathbb S}}             
\newcommand{\T}{{\mathbb T}}               
\newcommand{\D}{{\mathbb D}}               

\newcommand{\half}{\frac{1}{2}}
\renewcommand{\Re}{\mbox{\rm Re}}
\renewcommand{\Im}{\mbox{\rm Im}}
\newcommand{\sprod}[2]{\left\langle#1,#2\right\rangle}
\newcommand{\inner}[2]{\left\langle#1,#2\right\rangle_2}

\newcommand{\e}{\epsilon}
\newcommand{\w}{\omega}
\newcommand{\f}{\frac}

\newcommand{\ep}{\varepsilon}
\newcommand{\al}{\alpha}
\newcommand{\h}{\hbar}
\renewcommand{\tilde}{\widetilde}

\thispagestyle{empty}

\begin{abstract}
In these lectures we consider some topics of Quantum Field Theory
in Curved Space. In the first one particle creation in curved
space is studied from a mathematical point of view, especially,
particle production  at a given time using the so called
"instantaneous diagonalization method". As a first application we study
particle production in a no-oscillating model where re-heating may
be explained from gravitational particle creation. In the second one   we
re-calculate, with all the  mathematical details, particle
production in the Starobinsky model. Particle production by strong
electromagnetic fields (Schwinger's effect) and particle
production by moving mirrors simulating black hole collapse are also studied. In the second lecture
we calculate the re-normalized two-point function using the
adiabatic regularization. The conformally and minimally coupled
cases are considered for a scalar massive and massless field.
We reproduce previous results in a rigorous mathematical
form and clarify some empirical approximations and bounds. The re-normalized stress tensor is also calculated in
several situations. Finally, in last lecture quantum correction
due to a massless fields conformally coupled
 with gravity  are considered in order to study the avoidance of
singularities that appear in the flat Friedmann-Robertson-Walker
(FRW) model. It is assumed that the universe contains a barotropic
perfect fluid with state equation $p=\omega\rho$  (being $\rho$
the energy density and $p$ the pressure). The dynamics of the
model is studied for all values of the parameter $\omega$, and
also for all values of the two parameters, that we will call
$\alpha$ and $\beta$, provided by the quantum corrections. We will
see that only the case $\alpha>0$ could avoid the singularities.
Then  when $\omega>-1$, in order to obtain an expanding Friedmann
universe at late times (only a one-parameter family of solutions,
no a general solution, has this behavior at late times),
 the initial conditions of the no-singular solutions at
early times must be very fine tuned. These no-singular solutions
are: a general solution
 (a two-parameter family) leaving the contracting de Sitter phase, and a one-parameter
 family leaving the contracting Friedmann stage.
 On the other hand for $\omega<-1$ (phantom field), the
problem of the avoidance of  singularities is more involved
because if one considers an expanding Friedmann stage  at early
times, then instead of fine tune the initial conditions one also
 has to fine tune the parameters $\alpha$ and $\beta$ to obtain
a behavior without future singularities, because only a
one-parameter family of solutions  follows a contracting Friedmann
phase at late times, and only a particular  solution behaves like
a contracting de Sitter universe. The rest of solutions have
future singularities.
\end{abstract}


\maketitle

\vspace{-6.5mm}

\hspace*{10mm}  {\footnotesize Keywords: 
particle production,  Hamiltonian
diagonalization, vacuum fluctuations, singularity avoidance.}

\tableofcontents

\section{Introduction}
In these lectures we'll try to give a self-consistent presentation
of the quantum field theory in curved space and also to show some
of its applications. In the first lecture  we will  give a
mathematical presentation of the subject and we will re-derive,
with all the details, some of its applications to the theory of
gravitational re-heating. Our presentation is no-standard, in the
sense that we start explaining the ``adiabatic vacuum
prescription'' introduced by Parker in his thesis
\cite{p68,p69,p70}, and the
``instantaneous diagonalization
method'', introduced in Russian literature at the beginning of
70's, based in the idea that the number of created particles in a
given mode at a given time is the energy of the mode at this time
divided by the energy of a single particle in that mode
\cite{glm74,gmm76,gmm94,zs72}. As an instructive example we calculate
particle creation in the flat Friedmann-Robertson-Walker (FRW)
chart of the de Sitter space where one can sees the
difference between both prescriptions. After that, we introduce
the ``in'' and ``out'' states in asymptotically flat spaces where
particle creation can be defined in the standard way
\cite{f97,j04,w06,bd82,f85,pt09}. As an application, we
re-calculate the gravitational particle production in a transition
from the de Sitter phase to the radiation-dominated one
\cite{f87}, and we also discuss the problem of a second
inflationary stage related with the back-reaction
\cite{fkl99,fkl99a,fmsvv09}. This example is important because it
describes approximately the inflationary phase followed by a
transition to a radiation dominated universe, and the particle
production process may be used to explain the pre-heating in
inflationary non-oscillatory models \cite{f87,s93,pv99}. Finally
we study, in a great detail, particle production in the
Starobinsky model \cite{s80,v85,op06} because we believe that
there isn't a clear explanation of particle creation in this
model. In this case, in order to obtain the well-known results,
firstly one must disregards the power-law expansion of the universe
and only retains the oscillating behavior of the scale factor at
late times, secondly one has to assume that the energy density of
the created particles is a well-defined quantity in this model and
then one has  to choose a particular form of it, and finally one has
to assume that one kind of particles, named scalarons, are the
responsible for the late time behavior of the universe, and also
that is the decay of these scalarons
what produce particles.

Particle production by strong electromagnetic fields \cite{gmr85}
are also studied, and Schwinger's formula \cite{s51} that
calculates the probability that the vacuum state remains unchanged
in the presence of a constant electric field is deduced, in an
elementary but not at all mathematically correct way, using
standard methods of Quantum Field Theory. More precisely, it is
deduced calculating the Bogoliubov coefficients for every mode via
the W.K.B. method in the complex plane. At the end a rigorous demonstration is
outlined.

The last part of  this first lecture is devoted to the study of
particle production by moving mirrors. Our interest is
concentrated in trajectories that simulate the black body
collapse, and to get the same kind of results obtained by Hawking
in \cite{haw75}, i.e., to obtain the black body spectrum. This
occurs for perfect reflecting mirrors, but we'll show that for
semi-transparent moving mirrors  the radiation spectrum is a bit
different.

In the second lecture the vacuum quantum fluctuations are studied.
We calculate the re-normalized two-point function subtracting
adiabatic modes up to order two. We do the calculation  for
conformally coupled fields and also for minimally coupled ones
re-obtaining, in a consistent way, all the early well-known results.
Calculating  the two-point function is very important in the
context of inflation, for example,
the eternal inflation phenomenon is manifested in some inflationary model
(new
inflationary universe \cite{l82,v83, hm83}, chaotic inflation
\cite{l83}), i.e.,
the large-scale quantum fluctuations of the inflaton field termed
by its two-point function lead to 
a process of infinitely
self-reproducing inflationary mini-universes \cite{l85}. Studying   the back-reaction of  particles produced in the
pre-heating phase  also requires  the two-point function of the inflaton
field and the two-point function of the light particles involved in such a process \cite{f99,an08,f09}.

 The mean
problem with the two-point function is that it is ultra-violet
divergent and requires re-normalization. The simplest method for
obtaining  divergence-free expressions is the adiabatic regularization
based on  subtracting  some  generalized WKB modes \cite{b78}
that  have the same behavior at large frequencies as the exact
modes;  the divergent terms then cancel.
But the procedures for calculating 
 the re-normalized two-point
function differed somewhat  in early work. The authors assumed some not 
fully justified frequency cut-off or made unjustified approximations in
order to obtain finite quantities \cite{l82,s82,vf82,v83a}.

Our task in  this lecture  is to  matematically clarify the features appearing in those works.
Firstly, the massless case is studied (conformally coupled and
minimally coupled case), reproducing all the previous results in full detail. After this, we study a
massive field in the de Sitter phase where it's assumed that its
mass is smaller than the Hubble parameter (this is  typical in the inflationary models \cite{l85}).
Here  we derive the two-point function at late time in full
 detail and accurately demonstrate  the mean
formula obtained in \cite{l82}. We finish this lecture  reviewing
some important results about  the stress
tensor re-nomalization which will be used in last lecture.

In last lecture we study the avoidance of cosmological
singularities if one takes into account the vacuum corrections due
to a massless conformally coupled field.

It's well-known that  the classical solutions of the general
relativity for a Friedmann-Robertson-Walker (FRW) model contain,
in general, singularities (Big Bang, Big Rip, future sudden
singularities), this means that near these singularities the
space-time curvature is arbitrarily large. Then, for curvatures on
the order of the Planck length, quantum effects have to be taken
into account. These quantum effects, can violate the so-called
energy conditions \cite{m-pv99}, and consequently they can  modify
drastically the classical solution. For this reason, it is
possible that quantum effects avoid the classical singularities
\cite{d77,pf73}.

We consider the quantum effects produced by massless fields
conformally coupled with gravity. This is an special case where,
for a flat FRW universe, the quantum vacuum stress tensor, that
depends on two regularization parameters, that we call $\alpha$
and $\beta$, can be calculated explicitly. Then due to the trace
anomaly and the equation of conservation, one easily calculates
the vacuum energy density that contributes to the modified
Friedmann equation. This equation cannot be analytically
integrated, but a qualitative phase-space study can be performed.
This is the main objective of this lecture.

First, we introduce the quantum effects and
 write the modified Friedmann equation that depends on the
parameters $\alpha$ and $\beta$, which we'll assume that can take
all possible values. After that, we study the simplest case, i.e.
$\alpha=0$, in this case the modified Friedmann equation becomes a
first order differential equation and can be integrated. Our
conclusion, in that case, is that the singularities are not
avoided. Another simple case corresponds to the case of an empty
universe (it doesn't contain any barotropic fluid, only quantum
effects are taken into account). An special case ($\alpha<0$,
$\beta<0$) is the Starobinsky model \cite{s80}. We'll see that, in
that case, all solutions contain singularities, except  when
$\beta<0$, where it appears an unstable de Sitter solution, and an
unstable solution that connects the de Sitter solution with the
point $H=\dot{H}=0$ (being $H$ the Hubble parameter). Finally, we
study the general case, i.e., a universe filled by a barotropic
perfect fluid with state equation $p=\omega\rho$. The only case
where no-singular solutions may appear is when $\alpha>0$ and
$\beta<0$. Then taking the same point of view as
 \cite{aw86,wa85}, we show that when $\omega>-1$, the no-singular
early time behaviors that can lead, at late times, to the
Friedmann expanding stage, are a contracting de Sitter phase and a
contracting Friedmann phase. However their initial conditions can
be very fine tuned in order to match with the expanding Friedmann
stage. On the other hand, for $\omega<-1$ the late time behavior
of no-singular solutions that  come from the expanding Friedmann
stage at early time, are the contracting Friedmann phase and the
contracting de Sitter one. In this case, instead of fine tune the
initial conditions, one also has to fine tune the parameters
$\alpha$ and $\beta$ in order to obtain no-singular solutions. But
in both cases, the no-singular solutions are unstable in the sense
that an small perturbation leads them to a singular behavior.

The units used in these lectures are $c=\hbar=1$.

\section{Particle creation by classical fields}
Particle creation by gravitational fields is studied in this section. The theory developed is applied to
a non-oscillating inflationary model and to the Starobinsky one.
\subsection{Graviational particle production}

\subsubsection{Quantum fields in curved space-time: General Theory}

It's well known that
the Lagrangian density of a scalar field is \cite{bd82} ${\mathcal
L}=\frac{1}{2}(\partial_{\mu}\phi\partial^{\mu}\phi -m^2\phi^2-\xi
R\phi^2)$, and its  corresponding  Klein-Gordon equation is given by
\begin{eqnarray}\label{a1}
 (-\nabla_{\mu}\nabla^ {\mu}+ m^2+\xi R)\phi=0,
\end{eqnarray}
where $\xi$ is the coupling constant and $R$ is the scalar
curvature. If one considers the flat FRW metric
$ds^2=-dt^2+C(t)d{\bf x^2}= C(\eta)(-d\eta^2+d{\bf x^2})$ (being
$\eta$ the conformal time), the modes of the form $\phi_{\bf k}({\bf
x},\eta)\equiv (2\pi)^{-3/2}C^{-1/2}(\eta)e^{i{\bf kx}} \chi_{\bf
k}(\eta)$ will satisfy the equation
\begin{eqnarray}\label{a2}
\chi''_{\bf k}(\eta)+\Omega_{\bf k}^2(\eta)\chi_{\bf k}(\eta)=0,
\end{eqnarray}
where we have introduced the notation $\Omega_{\bf k}^2(\eta)\equiv
\omega_{\bf k}^2(\eta)+(\xi-1/6)C(\eta)R(\eta)$, with $\omega_{\bf
k}^2(\eta)=m^2C(\eta)+|{\bf k}|^2$ and
$R(\eta)=3\left(\frac{C''}{C^2}-\frac{1}{2}\frac{C'^2}{C^3}\right)=6\frac{a''}{a^3}$
(being $C\equiv a^2$).

Firstly, we are interested in the conformally coupled case, i.e., in
the case $\xi=1/6$, where equation (\ref{a2}) becomes
\begin{eqnarray}\label{a3}
\chi''_{\bf k}(\eta)+\omega_{\bf k}^2(\eta)\chi_{\bf k}(\eta)=0.
\end{eqnarray}

This is the equation of a set of no-interacting harmonic
oscillators, so to built a quantum theory, we can consider the
following Hamiltonian operator $\hat{\mathcal H}(\eta)\equiv
\frac{1}{2}\left(\hat{\sqcap}^2+\omega^2(\eta)\hat{\phi}^2\right)-\frac{1}{2}\omega(\eta)$
corresponding to a single harmonic oscillator.
 In  Heisenberg picture the operators
 $\hat{\sqcap}$ and $\hat{\phi}$ satisfy the equations ${\hat{\phi}}'=\hat{\sqcap}$ and
${\hat{\sqcap}}'=-\omega^2\hat{\phi}$, that is,  $\hat{\phi}$
satisfy the Klein-Gordon equation
${\hat{\phi}}''+\omega^2\hat{\phi}=0$, and thus,  one can writes:
\begin{eqnarray}\label{a4}
\left(\begin{array}{c}\hat{\phi}(\eta)\\
\hat{\sqcap}(\eta)
      \end{array}\right)=
\left(\begin{array}{c}\chi(\eta)\\
\chi'(\eta)
      \end{array}\right)
\hat{A}_{\chi}+
\left(\begin{array}{c}\chi^*(\eta)\\
\chi'^*(\eta),
      \end{array}\right)\hat{A}^{\dagger}_{\chi},
\end{eqnarray}
where the mode function $\chi$ is a solution of the Klein-Gordon
equation and $\hat{A}_{\chi}$ is, in that picture, a constant
operator  that we will call the "annihilation operator relative to
the mode $\chi$".

From the commutation relation $[\hat{\sqcap},\hat{\phi}]=-i$, one
can  deduces that $\chi$ must satisfies the relation
$\chi'\chi^*-\chi'^*\chi=-i$, which means that the annihilation
operator is given by
\begin{eqnarray}\label{a5}
\hat{A}_{\chi}=
-{i}\left(\chi'^*(\eta)\hat{\phi}(\eta)-\chi^*(\eta)\hat{\sqcap}(\eta)\right).
\end{eqnarray}

Once this operator has been introduced one  can defines  the "vacuum
state relative to the mode $\chi$", namely $|0;\chi\rangle$, as the
quantum state that satisfies $\hat{A}_{\chi}|0;\chi\rangle=0$. It's
clear from this definition that there isn't particle production at
any time, because
$\langle\chi;0|\hat{A}_{\chi}^{\dagger}\hat{A}_{\chi}|0;\chi\rangle=0$
 all the  time \cite{pf74}.
However, this definition depends on the choice of the mode $\chi$.
Effectively, if one chooses two different mode functions, namely
$\chi_1$ and $\chi_2$, since
$\hat{A}_{\chi_1}=\alpha_{1,2}\hat{A}_{\chi_2}+\beta_{1,2}^*\hat{A}_{\chi_2}^{\dagger}$
with $\alpha_{1,2}=-{i}{\mathcal W}[\chi_2;\chi_1^*]$ and
$\beta_{1,2}^*=-{i}{\mathcal W}[\chi_2^*;\chi_1^*]$ (where
${\mathcal W}$ denotes the Wronskian), then an observer in the
$|0;\chi_2\rangle$ vacuum state can observes $\chi_1$-particles
because one has ${\mathcal
N}_{1,2}\equiv\langle\chi_2;0|\hat{A}_{\chi_1}^{\dagger}\hat{A}_{\chi_1}|0;\chi_2\rangle=|\beta_{1,2}|^2.$

In this way, if one considers the family of solutions to the
Klein-Gordon equation, namely $\chi_{\eta'}(\eta)$, defined by the
initial condition
\begin{eqnarray}\label{a6}
\chi_{\eta'}(\eta')\equiv f(\eta');\quad \chi'_{\eta'}(\eta')\equiv
g(\eta'), \quad \mbox{where $f$ and $g$ are some arbitrary
functions},
\end{eqnarray}
one can calculates the number of $\chi_{\tau}$-particles detected by
an observer in the $|0;\chi_{\tau'}\rangle$ vacuum state, that is,
the number of produced particles at time $\tau$ from the vacuum
state at time $\tau'$,  with the formula
\begin{eqnarray}\label{a7}
{\mathcal
N}(\tau;\tau')\equiv\langle\chi_{\tau'};0_H|\hat{A}_{\chi_{\tau},H}^{\dagger}\hat{A}_{\chi_{\tau},H}|0_H;\chi_{\tau'}\rangle=|\beta(\tau;\tau')|^2,
\quad\mbox{with}\quad\beta(\tau;\tau')={i}{\mathcal
W}[\chi_{\tau'};\chi_{\tau}].
\end{eqnarray}

Note that, different families of solutions give rise to different
definitions of the vacuum state. For example, to define the {\bf
adiabatic vacuum modes} first we consider the
$\epsilon$-Klein-Gordon equation $\epsilon
\nu''+\omega^2(\eta)\nu=0$, where here, $\epsilon$ is a
dimensionless parameter that one shall set $\epsilon=1$ at the end
of the calculations. At order "$n$", a WKB solution of the
Klein-Gordon equation is (see for details \cite{wi05}):
\begin{eqnarray}\label{a8}
\chi_{n;WKB}(\tau;\epsilon)\equiv
\sqrt{\frac{1}{2W_n(\tau;\epsilon)}}e^{-\frac{i}{\epsilon}\int^{\tau}W_n(\eta;\epsilon)d\eta},
\end{eqnarray}
where $W_0=\omega$ and
\begin{eqnarray}\label{a9}
W_n=\mbox{terms until order $\epsilon^{2n}$ of}\left(\sqrt{
\omega^2-\epsilon^2\left[\frac{1}{2}\frac{W''_{n-1}}{W_{n-1}}-\frac{3}{4}\frac{(W'_{n-1})^2}{W^2_{n-1}}\right] }\right).
\end{eqnarray}

Once one has introduced the WKB solutions,  the adiabatic vacuum at
order $n$ is defined through the family  ${\chi}_{n;\eta'}(\eta)$
that satisfy the initial condition
\begin{eqnarray}\label{a10}
{\chi}_{n;\eta'}(\eta')=\chi_{n;WKB}(\eta';\epsilon=1);\quad
{\chi}'_{n;\eta'}(\eta')=\chi'_{n;WKB}(\eta';\epsilon=1).
\end{eqnarray}

From this definition, at the $n$ order, the $\beta$-Bogoliubov
coefficient is given by $\beta_n(\tau;\tau')=i{\mathcal
W}[\chi_{n;\tau'};\chi_{n;\tau}]$, and the number of produced
particles at time $\tau$ can be calculated from the formula
\begin{eqnarray}\label{a11}
{\mathcal N}_{n}(\tau;\tau')\equiv |\beta_{n}(\tau;\tau')|^2
\nonumber=&& \frac{1}{
W_n(\tau)}\left(\frac{1}{2}[|{\chi}'_{n;\tau'}(\tau)|^2+W_n^2(\tau)|{\chi}_{n;\tau'}(\tau)|^2]-
\frac{1}{2} W_n(\tau)\right)
\\&&
 +\frac{W_n'^2(\tau)}{8
W_n^3(\tau)}|{\chi}_{n;\tau'}(\tau)|^2 +\frac{W_n'(\tau)}{4
W_n^2(\tau)}({\chi}'^*_{n;\tau'}(\tau){\chi}_{n;\tau'}(\tau)+
{\chi}'_{n;\tau'}(\tau){\chi}^*_{n;\tau'}(\tau)),
\end{eqnarray}
where $W_n(\tau)\equiv W_n(\tau;\epsilon=1)$.

 Another important family  of solutions  to the
Klein-Gordon equation, namely ${\chi}_{diag;\eta'}(\eta)$, is given
by the initial condition
\begin{eqnarray}\label{a12}
{\chi}_{diag;\eta'}(\eta')=\chi_{0;WKB}(\eta';\epsilon=1);\quad
{\chi}'_{diag;\eta'}(\eta')=-i\omega(\eta')\chi_{0;WKB}(\eta';\epsilon=1).
\end{eqnarray}

This family defines the so-called {\bf instantaneous Hamiltonian
diagonalization method}, and their  Bogoliubov coefficients
\begin{eqnarray}\label{a13}
\alpha_{diag}(\tau;\tau')=-{i}{\mathcal
W}[\chi_{diag;\tau'};\chi^*_{diag;\tau}],\quad
\beta_{diag}(\tau;\tau')={i}{\mathcal
W}[\chi_{diag;\tau'};\chi_{diag;\tau}],
\end{eqnarray}
can be calculated as follows: Writing
${\chi}_{diag;\tau'}(\eta)=\alpha_{diag}(\tau;\tau'){\chi}_{diag;\tau}(\eta)
+\beta_{diag}(\tau;\tau'){\chi}^*_{diag;\tau}(\eta)$,  at
$\eta=\tau$ one gets the system
\begin{eqnarray}\label{a14}
\left\{\begin{array}{ccc}
{\chi}_{diag;\tau'}(\tau)&=&\alpha_{diag}(\tau;\tau')\chi_{0;WKB}(\tau;\epsilon=1)
+\beta_{diag}(\tau;\tau')\chi_{0;WKB}^*(\tau;\epsilon=1)\\
{\chi}'_{diag;\tau'}(\tau)&=&-i\omega(\tau)\left(\alpha_{diag}(\tau;\tau')\chi_{0;WKB}(\tau;\epsilon=1)
-\beta_{diag}(\tau;\tau')\chi_{0;WKB}^*(\tau;\epsilon=1)\right),
       \end{array}\right.
\end{eqnarray}
which can be used to obtain the interesting formula
\begin{eqnarray}\label{a15}
{\mathcal N}_{diag}(\tau;\tau')\equiv|\beta_{diag}(\tau;\tau')|^2=
\frac{1}{\omega(\tau)}\left(\frac{1}{2}[|\chi'_{diag;\tau'}(\tau)|^2+\omega^2(\tau)|\chi_{diag;\tau'}(\tau)|^2]-\frac{1}{2}\omega(\tau)\right),
\end{eqnarray}
which shows that this quantity is the energy at time $\tau$ of the
mode $\chi_{diag;\tau'}$ divided by the energy, at time $\tau$, of a
single particle.

Another way to obtain the Bogoliubov coefficients can be done from
the system (\ref{a14}) if one takes into account that the mode
function ${\chi}_{diag;\tau'}(\tau)$ satisfy the equation
${\chi}''_{diag;\tau'}(\tau)+\omega^2(\tau){\chi}_{diag;\tau'}(\tau)=0$.
 As a function of
the variable $\tau$, one has
\begin{eqnarray}\label{a16}
\left\{\begin{array}{ccc}
\alpha'_{diag}(\tau;\tau')&=& {\omega'(\tau)}\left(\nu^*_{0;WKB}(\tau;\epsilon=1)\right)^2\beta_{diag}(\tau;\tau')\\
\beta'_{diag}(\tau;\tau')&=&
{\omega'(\tau)}\left(\nu_{0;WKB}(\tau;\epsilon=1)\right)^2\alpha_{diag}(\tau;\tau').
       \end{array}\right.
\end{eqnarray}

This system can be solved by iteration, for example, if in the first
iteration one chooses $\alpha_{diag}(\tau;\tau')\cong 1$,  one will
arrive at formula
\begin{eqnarray}\label{a17}
\beta_{diag}(\tau;\tau')\cong
\int_{\tau'}^{\tau}\frac{\omega'(\eta)}{2\omega(\eta)}e^{-2i\int^{\eta}\omega(\eta')d\eta'}d\eta.
\end{eqnarray}

\begin{example}(Particle creation in the flat FRW chart of the de
Sitter space-time \cite{g94,he08})

In the de Sitter phase the scalar factor is given by
$a(\eta)=-1/(H\eta)$, with $-\infty<\eta<0$ (being $H$ the Hubble
parameter), and the frequency has the form
$\omega(\eta)=\sqrt{\omega_0^2+\frac{m^2}{H^2\eta^2}}$, where
$\omega_0$ is a constant. We are interested in the case $m\gg H$
which correspond to the adiabatic approximation (see below). An easy
calculation yields
\begin{eqnarray}\label{a18}
\chi_{0;WKB}(\eta;\epsilon=1)=\sqrt{\frac{1}{2\omega(\eta)}} e^{
i\omega_0\sqrt{\eta^2+ m^2/(H^2\omega_0^2)}} \left(\frac{|\eta|}
{\sqrt{\eta^2+ m^2/(H^2\omega_0^2)}+ m/(H\omega_0) }\right)^{im/H},
\end{eqnarray}
which shows,  when $\eta\rightarrow -\infty$, that
\begin{eqnarray}\label{a19}
\chi_{0;WKB}(\eta;\epsilon=1)\rightarrow
\sqrt{\frac{1}{2\omega_0}}e^{-i\omega_0\eta};\quad
\chi_{0;WKB}'(\eta;\epsilon=1)\rightarrow
-i\omega_0\nu_{0;WKB}(\eta;\epsilon=1).
\end{eqnarray}

The  mode solution that satisfy the initial condition (\ref{a19}) is
given in terms of the Hankel functions \cite{bd82}
\begin{eqnarray}\label{a20}
\chi(\eta)=C\sqrt{\frac{\pi\eta}{4}}H^{(2)}_{\mu}(\omega_0\eta),
\end{eqnarray}
 with $\mu\equiv\sqrt{\frac{1}{4}-\frac{m^2}{H^2}}\cong im/H$,\quad $C\equiv
e^{-i(\frac{\pi\mu}{2}+\frac{\pi}{4})}\cong e^{\pi
m/(2H)}e^{-i\frac{\pi}{4}}$ \quad and \quad $\eta=e^{-i\pi}|\eta|$.

 Using the asymptotic form of the Hankel functions at late times, i.e., when
$|\eta|\omega_0\ll 1$ \cite{as72}
\begin{eqnarray}\label{a21}
\chi(\eta)\cong -C\sqrt{\frac{\eta}{4\pi}}\frac{H}{m}\left[ e^{-\pi
m/H}\Gamma(1-im/H)\left(\frac{\omega_0\eta}{2}\right)^{im/H}-\Gamma(1+im/H)
\left(\frac{\omega_0\eta}{2}\right)^{-im/H}\right],
\end{eqnarray}
an easy calculation provides that
\begin{eqnarray}\label{a22}
|\beta_{diag}(0;-\infty)|^2 \cong \frac{H^3}{32\pi m^3}\left|\Gamma\left(1-im/H\right)\right|^2
e^{\pi m/H}.
\end{eqnarray}

Using at this point \cite{as72} $|\Gamma(1+iy)|^2=\pi y/\sinh(\pi
y)$, we conclude that, when $|\eta|\omega_0\ll 1$, the number of
produced particles, using the diagonalization method, is given by
\cite{he08}
\begin{eqnarray}\label{a23}
{\mathcal N}_{diag}(0;-\infty) \cong  \frac{H^2}{16 m^2}.
\end{eqnarray}

However if we use the zero order (the other orders give the same
result) adiabatic vacuum modes,  the square of the
$\beta$-Bogoliubov coefficient will be given by
\begin{eqnarray}\label{a24}
|\beta_0(0;-\infty)|^2=
\left|\chi(0)\chi'_{0;WKB}(0;\epsilon=1)-\chi'(0)\chi_{0;WKB}(0;\epsilon=1)\right|^2.
\end{eqnarray}

Inserting (\ref{a21}), in this last formula, we obtain
\begin{eqnarray}\label{a25}{\mathcal N}_{0}(0;-\infty)
\cong \frac{H}{2\pi m}\left|\Gamma\left(1+im/H\right)\right|^2
e^{-\pi m/H}= \left(e^{2\pi m/H}-1\right)^{-1}.
\end{eqnarray}

This is the thermal spectrum obtained in the flat FRW chart of the de Sitter space-time
\cite{gh77,g94}.

\begin{remark}
The two methods give a different result because when
$|\eta|\omega_0\ll 1$, one has
\begin{eqnarray}\label{a26}
\chi'_{0,WKB}(\eta;\epsilon=1)
\cong-i\omega(\eta)\chi_{0,WKB}(\eta;\epsilon=1)+\frac{1}{2\eta}\chi_{0,WKB}(\eta;\epsilon=1)
\not= -i\omega(\eta)\chi_{0,WKB}(\eta;\epsilon=1),\end{eqnarray}
this is due to the fact that $\lim_{\eta\rightarrow
0^-}\omega(\eta)=\infty$, that is, at late time there is not a
well-defined "out" region (see below for a precise definition of the
``out'' region).
\end{remark}

\end{example}

\vspace{1cm}

In general (for arbitrary values of $\xi$), for a given set of modes
satisfying ${\mathcal W}[\chi_{\bf k}^*,\chi_{\bf k}]=-i$ one can
expands the quantum field, in the Heisenberg picture, as follows:
$\hat{\phi}({\bf x},\eta)=\sum_{\bf k}\hat{A}_{\chi_{{\bf
k}}}\phi_{\bf k}({\bf x},\eta)+ \hat{A}_{\chi_{{\bf
k}}}^{\dagger}\phi_{\bf k}^*({\bf x},\eta)$, then one can defines
the quantum vacuum state relative to the modes $\phi_{\bf k}({\bf
x},\eta)= (2\pi)^{-3/2}C^{-1/2}(\eta)e^{i{\bf kx}} \chi_{\bf
k}(\eta)$, namely $|0; \chi\rangle$, which must satisfies
$\hat{A}_{\chi_{{\bf k}}}|0; \chi\rangle=0$ for all values of ${\bf
k}$. However if one considers another set of modes, namely
$\tilde{\phi}_{\bf k}({\bf x},\eta)$, one also may  develops the
quantum field as $\hat{\phi}({\bf x},\eta)=\sum_{\bf
k}\hat{A}_{\tilde{\chi}_{{\bf k}}}\tilde{\phi}_{\bf k}({\bf
x},\eta)+ \hat{A}_{\tilde{\chi}_{{\bf
k}}}^{\dagger}\tilde{\phi}_{\bf k}^*({\bf x},\eta)$, where
$\tilde{\phi}_{\bf k}\equiv \alpha_{\bf k}{\phi}_{\bf k}+\beta_{\bf
k}{\phi}_{\bf k}^*$ and thus, $\hat{A}_{\tilde{\chi}_{{\bf
k}}}\equiv \alpha_{\bf k}^*\hat{A}_{{\chi}_{{\bf k}}}-\beta_{\bf
k}^*\hat{A}_{\chi_{{\bf k}}}^{\dagger}$ with $|\alpha_{\bf
k}|^2-|\beta_{\bf k}|^2=1$. The vacuum $|{0}; \tilde{\chi}\rangle$
relative to the modes $\tilde{\phi}_{\bf k}({\bf x},\eta)$, is
related with the other vacuum trough the relation
\begin{eqnarray}\label{a27}
 |{0};\tilde{\chi}\rangle=\prod_{\bf k}\exp\left\{\frac{1}{2}\frac{\beta^*_{\bf k}}{\alpha^*_{\bf k}}
(\hat{A}_{\chi_{{\bf k};\tau}}^{\dagger})^2\right\}|{0};\chi\rangle,
\end{eqnarray}
and the operator "number of particles in the mode ${\bf k}$" that
depends on the choice of the set of modes, for example, for the set
$\phi_{\bf k}({\bf x},\eta)$ is $\hat{\mathcal N}_{{\chi}_{{\bf
k}}}\equiv \hat{A}_{\chi_{{\bf k}}}^{\dagger}\hat{A}_{\chi_{{\bf
k}}}$ satisfies $\langle \chi; 0|\hat{\mathcal N}_{{\chi}_{{\bf
k}}}|0;\chi\rangle=0$, however one has $\langle \tilde{\chi};
0|\hat{\mathcal N}_{{\chi}_{{\bf
k}}}|0;\tilde{\chi}\rangle=|\beta_{\bf k}|^2$.

Once we have introduced these definitions, one can considers the
family of solutions to the Klein-Gordon equation, namely $\chi_{{\bf
k};\eta'}(\eta)$, defined by the initial condition
\begin{eqnarray}\label{a28}
\chi_{{\bf k};\eta'}(\eta')\equiv f_{\bf k}(\eta');\quad \chi'_{{\bf
k};\eta'}(\eta') \equiv g_{\bf k}(\eta'), \quad \mbox{where $f_{\bf
k}$ and $g_{\bf k}$ are some arbitrary functions},
\end{eqnarray}
to calculate the number density of $\chi_{\tau}$-particles per unit
volume detected by an observer in the $|0;\chi_{\tau'}\rangle$
vacuum state, that is, the number density of produced particles at time
$\tau$ from the vacuum state at time $\tau'$.  The general formula
is
\begin{eqnarray}\label{a29}
N(\tau;\tau')\equiv\frac{1}{(2\pi a)^3}\int d^3{\bf k}
\langle\chi_{\tau'};0| \hat{A}_{\chi_{{\bf
k};\tau}}^{\dagger}\hat{A}_{\chi_{{\bf
k};\tau}}|0;\chi_{\tau'}\rangle=\frac{1}{(2\pi a)^3}\int d^3{\bf
k}|\beta_{\bf k}(\tau;\tau')|^2,
\end{eqnarray}
with $\beta_{\bf k}(\tau;\tau')={i}{\mathcal
W}[\chi_{{\bf k};\tau'};\chi_{{\bf k};\tau}]$.

\begin{remark}
In the conformally coupled case one also can defines the energy density of produced
particles at time $\tau$ from the vacuum state at time $\tau'$ as follows
$\rho(\tau;\tau')\equiv\frac{1}{(2\pi a)^3a}\int d^3{\bf
k}\omega_{\bf k}(\tau)|\beta_{\bf k}(\tau;\tau')|^2$.
\end{remark}

\begin{example}(Particle production in the adiabatic approximation)

This approximation is based in the assumption $\Omega'_{\bf k}\ll
\Omega^2_{\bf k}$. In the conformally coupled case this assumption
becomes $\omega'_{\bf k}\ll \omega^2_{\bf k}$ and it is always
satisfied when $H\ll m$. In that case one has
\begin{eqnarray}\label{a30}
 \chi_{{\bf k},diag;\tau'}(\tau)\cong \chi_{{\bf k}, 0;WKB}(\tau;\epsilon=1)=
\sqrt{\frac{1}{2\omega_{\bf
k}(\tau)}}e^{-{i}\int_{\tau'}^{\tau}\omega_{\bf k}(\eta)d\eta},
\end{eqnarray}
and  when $\omega_{\bf k}'(\tau')=0$ one can  inserts this
expression in formula (\ref{a15}) to obtain
 \begin{eqnarray}\label{a31}
|\beta_{{\bf k},diag}(\tau;\tau')|^2\cong \frac{\omega_{\bf
k}'^2(\tau)}{16\omega_{\bf k}^4(\tau)} =\frac{m^4
C'^2(\tau)}{64\omega_{\bf k}^6(\tau)},
\end{eqnarray}
which helps us to  conclude that, using the instantaneous
diagonalization method,
 the number density of created particles per unit volume is given by
 \cite{gmm76}
\begin{eqnarray}\label{a32}
 N_{diag}(\tau;\tau')\cong \frac{mH^2}{32\pi^2}\int_0^{\infty}\frac{x^2}{(x^2+1)^3}dx=
\frac{mH^2}{512\pi},
\end{eqnarray}
and their energy density is
\begin{eqnarray}\label{a33}
 \rho_{diag}(\tau;\tau')\cong \frac{m^2H^2}{32\pi^2}\int_0^{\infty}\frac{x^2}{(x^2+1)^{5/2}}dx=
\frac{m^2H^2}{96\pi}.
\end{eqnarray}
\begin{remark}
 Note that   formula (\ref{a23}) can be  easily obtained from formula  (\ref{a31}) applied to the de Sitter
phase.
\end{remark}
\end{example}

To finish 
 we consider and asymptotically flat FRW space-time, that is, we
assume that $\lim_{\eta\rightarrow\pm\infty} C(\eta)=C_{\pm}$, and
we take the following set of modes $\phi_{in, \bf k}({\bf
x},\eta)\equiv (2\pi)^{-3/2}C^{-1/2}(\eta)e^{i{\bf kx}}
\frac{e^{-i\omega_{-,\bf k}\eta}}{\sqrt{2\omega_{-,\bf k}}}$ when
$\eta\rightarrow -\infty$, and the the set $\phi_{out, \bf k}({\bf
x},\eta)\equiv (2\pi)^{-3/2}C^{-1/2}(\eta)e^{i{\bf kx}}
\frac{e^{-i\omega_{+,\bf k}\eta}}{\sqrt{2\omega_{+,\bf k}}}$ when
$\eta\rightarrow \infty$ (being $\omega_{\pm, \bf
k}=\sqrt{m^2C_{\pm}+|{\bf k}|^2}$). In this context we can define
the ``in'' and ``out'' vacuum states, namely $|0_{in}\rangle$ and
$|0_{out}\rangle$, and the ``in'' and ``out'' annihilation operators
$\hat{A}_{in,\bf k}$ and $\hat{A}_{out,\bf k}$. Then, the average
number of produced pairs, at late times, in the  $\bf k$ mode, is
given by $\langle 0_{in}|\hat{\mathcal N}_{out,\bf k}|
0_{in}\rangle=|\beta_{\bf k}|^2$, where $\hat{\mathcal N}_{out,\bf
k}=\hat{A}_{out,\bf k}^{\dagger}\hat{A}_{out,\bf k}$ is the operator
number of ``out'' particles, and the beta Bogoliubov coefficient is
obtained trough the relation $\phi_{in,\bf k}\equiv \alpha_{\bf
k}{\phi}_{out,\bf k}+\beta_{\bf k}{\phi}_{out,\bf k}^*$.

In that context, the number density of created particle per unit
volume, at late times, is given by
\begin{eqnarray}\label{a34}
 N=\frac{1}{(2\pi a)^3}\int d^3{\bf k}|\beta_{\bf k}|^2,
\end{eqnarray}
 and their energy density by \cite{bd79}
\begin{eqnarray}\label{a35}
 \rho=\frac{1}{(2\pi a)^3a}\int d^3{\bf k}\omega_{+,\bf k}|\beta_{\bf k}|^2.
\end{eqnarray}

In general, it is impossible to solve the mode equation (\ref{a2})
but one can rewrite
 this differential equation
in an integral one as follows \cite{bd79,zs77}:
\begin{eqnarray}\label{a36}
 \chi_{\bf k}(\eta)=\frac{e^{-i\omega_{-,\bf k}\eta}}{\sqrt{2\omega_{-,\bf k}}}+\frac{1}{\omega_{-,\bf k}}
\int_{-\infty}^{\eta}V_{\bf k}(\eta')\sin(\omega_{-,\bf k}(\eta-\eta'))\chi_{\bf k}(\eta')d\eta',
\end{eqnarray}
where $V_{\bf k}(\eta)=\omega_{-,\bf k}^2-\Omega_{\bf k}^2(\eta)$.

Applying Picard's method to lowest order, i.e, replacing $\chi_{\bf
k}(\eta')$ by $\frac{e^{-i\omega_{-,\bf
k}\eta'}}{\sqrt{2\omega_{-,\bf k}}}$ one obtains the following
approximation that works very well for  massless  nearly conformally
coupled fields
\begin{eqnarray}\label{a37}
\alpha_{\bf k}\cong 1+\frac{i}{2\omega_{-,\bf k}}\int_{\R}V_{\bf
k}(\eta)d\eta \quad \beta_{\bf k}\cong -\frac{i}{2\omega_{-,\bf
k}}\int_{\R} e^{-2i\omega_{-,\bf k}\eta} V_{\bf k}(\eta)d\eta
\end{eqnarray}

Note that, one will have to assume $\lim_{\eta\rightarrow
\pm\infty}V_{\bf k}(\eta)=0$ if one wants
 well defined Bogoliubov coefficients. This always happens in the massless case, and
using Plancherel's theorem is not difficult, in the massless case, to prove that
\begin{eqnarray}\label{a38}
 N=\frac{(\xi-1/6)^2}{16\pi a^3}\int_{\R}a^4(\eta)R^2(\eta)d\eta.
\end{eqnarray}

\subsubsection{Particle production in the transition from de Sitter phase
to a radiation dominated universe}

Consider the following scale factor \cite{f87}
\begin{eqnarray}\label{a39}
a(\eta)=\left\{\begin{array}{ccc} -\frac{1}{H\eta}& \mbox{for}&
\eta\leq\eta_0\\
H(\eta-\eta_0)-\frac{1}{H\eta_0}& \mbox{for} &\eta_0\leq\eta,
\end{array}\right.
\end{eqnarray}
where $\eta_0<0$ is the time when the sudden transition occurs.

This example is interesting because in the radiation phase massless
particles cannot be produced, so in that phase, the number density of
created particles and their energy density are well defined
quantities. Moreover, it describes approximately the inflationary
phase followed by a transition to a radiation dominated universe,
and the obtained result can be used to explain the reheating process
of the universe after inflation in some no-oscillatory models
\cite{pv99}.

 First, we wil consider  massless  nearly conformally coupled particles. In the de Sitter phase, one has
$\Omega_{\bf k}^2(\eta)=|{\bf k}|^2+(12\xi-2)/\eta^2$, and in the
radiation one $\Omega_{\bf k}^2(\eta)=|{\bf k}|^2$. Using formula
(\ref{a37}) one gets
\begin{eqnarray}\label{a40}
 \beta_{\bf k}\cong \frac{i}{|{\bf k}|}\int_{-\infty}^{\eta_0}
e^{-2i|{\bf k}|\eta}\frac{6\xi-1}{\eta^2} d\eta,
\end{eqnarray} and
form formula (\ref{a38}) one easily obtains
\begin{eqnarray}\label{a41}
 N=-\frac{(6\xi-1)^2}{12\pi
 a^3\eta_0^3}=\frac{(6\xi-1)^2H^3}{12\pi}\left(\frac{a_0}{a}\right)^3,
\end{eqnarray}
where $a_0\equiv a(\eta_0)$.

It isn't difficult to show, from formula (\ref{a37}), that the
energy density diverges. However if one assumes that the transition
is not abrupt one obtains (see \cite{f87})
\begin{eqnarray}\label{a42}
 \rho\sim \frac{(6\xi-1)^2}{ a^4\eta_0^4}={(6\xi-1)^2H^4}\left(\frac{a_0}{a}\right)^4.
\end{eqnarray}

Now, we consider massless minimally coupled particles where the mode
functions that describe the vacuum state in the de Sitter phase are
given by
\begin{eqnarray}\label{a43}
 \chi_{\bf k}(\eta)=-\sqrt{\frac{\pi\eta}{4}}H^{(2)}_{3/2}(|{\bf k}|\eta)=\sqrt{\frac{1}{2|{\bf k}|}}e^{-i|{\bf
k}|\eta}\left(1+\frac{1}{i|{\bf k}|\eta}\right),
\end{eqnarray}
and in the radiation one, these modes have the form
\begin{eqnarray}\label{a44}
 \chi_{\bf k}(\eta)=\frac{1}{\sqrt{2|{\bf k}|}}\left(\alpha_{\bf k}e^{-i|{\bf k}|\eta}+
\beta_{\bf k}e^{i|{\bf k}|\eta}\right).
\end{eqnarray}

Matching at time $\eta=\eta_0$ one obtains $\beta_{\bf
k}=\frac{1}{2|{\bf k}|^2\eta_0^2}e^{-2i|{\bf k}|\eta_0}$ what
implies that the number density of produced particles is infrared
divergent, and their energy density is both, infrared and
ultraviolet divergent.

To eliminate the infrared divergency one can imagines that at very
early times the universe is in a radiation phase, then at a given
time, for example  $\eta=-H^{-1}$ ($t=0$), there is an abrupt
transition to the inflationary phase described approximately by a de
Sitter one (see for details \cite{fp77,vf82}). On the other hand, to
avoid the ultraviolet divergencies and can assumes that inflation
finishes with an smooth transition to the radiation phase \cite{f87}
because in that case modes with $|k|\gg |\eta|_0^{-1}$ have a very
small contribution. Anyway if one is only interested in production
of particles whose modes leave the Hubble horizon, that is, in modes
that satisfy $H<|{\bf k}|< |\eta_0|^{-1}$, one can uses the formula
$\beta_{\bf k}=\frac{1}{2|{\bf k}|^2\eta_0^2}e^{-2i|{\bf k}|\eta_0}$
to obtain
\begin{eqnarray}\label{a45}
 N=\frac{1}{8\pi^2a^3\eta_0^4}\int_{H}^{|\eta_0|^{-1}}|{\bf k}|^{-2}d|{\bf k}|=\frac{H^3}{8\pi^2}\left(\frac{a_0}{a}\right)^3\left(\frac{1}{H|\eta_0|}-1\right),
\end{eqnarray}and
\begin{eqnarray}\label{a46}
 \rho=\frac{1}{8\pi^2a^4\eta_0^4}\int_{H}^{|\eta_0|^{-1}}|{\bf k}|^{-1}d|{\bf
 k}|=\frac{H^4}{8\pi^2}\left(\frac{a_0}{a}\right)^4\ln\left(\frac{1}{H|\eta_0|}\right).
\end{eqnarray}

Now we consider the general massless case where the modes that
defines the vacuum state in the de Sitter phase are
\begin{eqnarray}\label{a47}
 \chi_{\bf k}(\eta)=C\sqrt{\frac{\pi\eta}{4}}H^{(2)}_{\nu}(|{\bf k}| \eta),
\end{eqnarray}
with $\nu\equiv\sqrt{\frac{9}{4}-12\xi}$ and $C\equiv
e^{-i(\frac{\pi\nu}{2}+\frac{\pi}{4})}$.

For modes that satisfy $|{\bf k}\eta_0|\ll 1$ one can uses the
asymptotic formula
\begin{eqnarray}\label{a48}
 H^{(2)}_{\nu}(z)\cong \frac{i}{\pi}\left(z/2\right)^{-\nu}\Gamma(\nu)-\frac{ie^{i\pi\nu}}{\sin(\pi\nu)}\left(z/2\right)^{\nu}
\frac{1}{\Gamma(\nu+1)}.
\end{eqnarray}

Then matching at point $\eta=\eta_0$ one obtains for $\nu>0$ (i.e.,
for $\xi<3/16$) \cite{dv96}
\begin{eqnarray}\label{a49}
 |\beta_{\bf k}|^2\cong \frac{1}{16\pi}\left(|{\bf
 k}||\eta_0|/2|\right)^{-2\nu-1}\Gamma^2(\nu)\left(1/2-\nu\right)^2.
\end{eqnarray}

In the opposite case $|{\bf k}\eta_0|\gg 1$, from the asymptotic
formula \cite{as72}
\begin{eqnarray}\label{a50}
 H^{(2)}_{\nu}(z)\cong C\sqrt{\frac{2}{\pi z}}
\left(1-i\frac{4\nu^2-1}{8z}\right)e^{-iz},
\end{eqnarray}
after matching at point $\eta=\eta_0$ one obtains
\begin{eqnarray}\label{a51}
 |\beta_{\bf k}|^2\cong \frac{1}{16}\frac{4\nu^2-1}{|{\bf k}|^2\eta_0^2}e^{-i|{\bf k}|\eta_0},
\end{eqnarray}
what means that the energy density is always ultraviolet divergent.

From these results one concludes that for $5/48<\xi<3/16$
($0<\nu<1$) there isn't infrared divergencies and thus the number
density of particles converges. Moreover, if one assumes
 that the transition is smooth then their energy density is also finite. On the other hand, when
$\xi\leq 5/48$ ($\nu\geq 1$) the number density is infrared
divergent (and for $\xi\leq 0$  ($\nu\geq 3/2$) their energy is also
infrared divergent). The solution to the avoidance of divergencies
in this last case is the same as for the minimally coupled case.
Then if one is only interested in
 modes that leave the Hubble Horizon,
 one can uses the approximation $|{\bf k}\eta_0|\ll 1$ to
obtain the formulae
\begin{eqnarray}\label{a52}
 N\cong \frac{4^{\nu}}{32\pi^3(\nu-1)}\Gamma^2(\nu)\left(1/2-\nu\right)^2{H^3}\left(\frac{a_0}{a}\right)^3
\left(\frac{1}{(H|\eta_0|)^{2(\nu-1)}}-1\right),
\end{eqnarray}
and
\begin{eqnarray}\label{a53}
 \rho\cong \frac{4^{\nu}}{64\pi^3(2\nu-3)}\Gamma^2(\nu)\left(1/2-\nu\right)^2{H^4}\left(\frac{a_0}{a}\right)^4
\left(\frac{1}{(H|\eta_0|)^{(2\nu-3)}}-1\right).
\end{eqnarray}

\begin{remark} Note that the case $\nu=1/2$ is the conformally
coupled case and thus there isn't particle production.
\end{remark}

To obtain  results for massive particles we will apply the
diagonalization method, more precisely, to calculate $|\beta_{{\bf
k},diag}(\eta_0;-\infty)|^2$ we have to use formula (\ref{a15})
with, and that is very important, $\omega_{\bf
k}(\eta_0)=\sqrt{|{\bf k}|^2+\frac{m^2}{H^2|\eta_0|^2}}$. Then for
light particles (particles with small mass compared with the Hubble
parameter, i.e., with $m\ll H$) no-conformally coupled one can uses
the formulae (\ref{a52}) and (\ref{a53}). However for conformally
coupled particles we've obtained, after a cumbersome calculation, in
the range $m/H\ll |{\bf k}\eta_0|\ll 1$
\begin{eqnarray}\label{a54}
|\beta_{{\bf k},diag}(\eta_0;-\infty)|^2\sim
\frac{m^2}{H^4\eta_0^4|{\bf k}|^4},
\end{eqnarray}
then, integrating over the modes that leave the Hubble horizon one
has
\begin{eqnarray}\label{a55}
\rho_{diag}(\eta_0)\sim
m^4\left(\frac{a_0}{a}\right)^4\ln\left(\frac{1}{H|\eta_0|}\right).
\end{eqnarray}

In the opposite case, that is, for $m\gg H$, from the example $2.1$
in the range $|{\bf k}\eta_0|\ll 1$, one gets
\begin{eqnarray}\label{a56}
|\beta_{{\bf k}, diag}(\eta_0;-\infty)|^2\sim \frac{H^2}{m^2},
\end{eqnarray}
and integrating over the modes that leave the Hubble horizon one has
\begin{eqnarray}\label{a57}
\rho_{diag}(\eta_0)\sim \frac{H^5}{m}\left(\frac{a_0}{a}\right)^4.
\end{eqnarray}

Finally it's important to remark that for light particles it is
well-known that its  energy density  during the de Sitter phase is
of the order $H^4$ (see equation (\ref{b179}) in lecture II), and  then if one takes a
inflationary model where the energy density of the inflaton field,
namely $\rho_v$, satisfies $\rho_v\ll m_p^4$ (being $m_p$ the Planck
mass), for example if it is at the grand unified theory scale
\cite{g04}, then from the Friedmann equation $H^2=\frac{8\pi
\rho_v}{3 m_p^2}$ one will deduce that $H^4\ll \rho_v$, that is, the
back-reaction effect will be negligible. However, when $\rho_v\sim
m_p$, in \cite{fkl99,fkl99a} the authors showed, in the minimally
coupled case, that  there is a new inflationary state (which makes
the results obtained above incorrect) driven by the field that
produce light particles, namely $\phi$, and not by the inflaton
field, namely $\Phi$, and thus instead of studying gravitational
particle production, one should study the mechanism of production of
particles of the field $\Phi$ by the oscillations of the field
$\phi$ due to the potential energy density  $\frac{1}{2}m^2\phi^2$.

\subsubsection{Particle production in the Starobinsky model}

In this model the universe emerges form the de Sitter phase and at
late times the scale factor is approximately given by (see for
example \cite{s80,v85}) $a(t)\cong t^{2/3}[1+\frac{2}{3Mt}\sin(Mt)]$
(in terms of the Hubble parameter
$H(t)=\frac{4}{3t}\cos^2\left(Mt/2\right)\left[1-\frac{\sin(Mt)}{Mt}\right]$
where $M$ is a constant that is related with the vacuum polarization
effect produced by massless conformally coupled particles. More
precisely, this model corresponds to an empty universe (it doesn't
contain any barotropic fluid), and only quantum effects due to
massless conformally coupled fields are taken into account
\cite{s80,v85,aw86,op06}. The energy density that contributes to the
Friedmann equation depends on two parameters \cite{d77,fhh79,a83},
one of them is $M$ and the other is called $H_0$ in \cite{v85}.
 Then, when these two parameters are positive there is an
unstable no-singular solution, that emerges from the de Sitter phase
with $H(t)=H_0$ and at late times  approaches to a matter dominated
universe with
$H(t)=\frac{4}{3t}\cos^2\left(Mt/2\right)\left[1-\frac{\sin(Mt)}{Mt}\right]$
(A more detailed description of this model is given in last lecture).

First at all we study the production of massless nearly conformally
coupled particles. To do this we'll follow Vilenkin's viewpoint
\cite{v85} (see also \cite{op06}). The idea is to calculate the rate
of particle per unit volume and per unit time, this concept was
introduced in \cite{zs77} but without any explanation. Here we'll
try to justify the results obtained in  \cite{zs77}. Note that to
calculate the particle rate per unit time one needs to calculate the
particle production at a given time and then to calculate its
derivative with respect to time, that is, one has to use formula
(\ref{a29}) with $\tau'=-\infty$. To perform this calculation we've
to choose a given family of mode solutions. From our point of view
the most natural choice is
\begin{eqnarray}\label{a58}
 \chi_{\bf k, \tau}(\eta)=\frac{e^{-i|{\bf k}|\eta}}{\sqrt{2|{\bf k}|}}+\frac{1}{|{\bf k}|}
\int_{\tau}^{\eta}V_{\bf k}(\eta')\sin(|{\bf
k}|(\eta-\eta'))\chi_{\bf k,\tau}(\eta')d\eta',
\end{eqnarray}
with $V_{\bf k}(\eta)=\omega_{-,\bf k}^2-\Omega_{\bf k}^2(\eta)=-(\xi-1/6)C(\eta)R(\eta)$.

Note that these modes satisfy
\begin{eqnarray}\label{a59}
 \chi_{\bf k, \tau}(\tau)=\frac{e^{-i|{\bf k}|\tau}}{\sqrt{2|{\bf k}|}} \quad
\mbox{and} \quad \chi_{\bf k, \tau}'(\tau)=-i|{\bf k}|\frac{e^{-i|{\bf k}|\tau}}{\sqrt{2|{\bf k}|}}.
\end{eqnarray}

Now applying Picard's method to lowest order one obtains
\begin{eqnarray}\label{a60}
 \beta_{\bf k}(\tau;-\infty)\cong -\frac{i}{2|{\bf
k}|}\int_{\R} \theta(\tau-\eta)e^{-2i|{\bf k}|\eta} V_{\bf k}(\eta)d\eta,
\end{eqnarray}
where we've introduced the Heaviside function $\theta$. Then from
Plancherel's theorem one obtains the following density of produced
particles at time $\tau$
\begin{eqnarray}\label{a61}
 N(\tau;-\infty)\cong \frac{(\xi-1/6)^2}{16\pi a^3}\int_{-\infty}^{\tau}a^4(\eta)R^2(\eta)d\eta.
\end{eqnarray}

To calculate the rate of particles per unit volume and unit time,
one has to multiply this last quantity by $a^3$  and to take the
derivative with respect the cosmic time and finally to divide by
$a^3$, the final result is
\begin{eqnarray}\label{a62}
 \frac{1}{a^3}\frac{d}{dt}(a^3N(\tau;-\infty))\cong \frac{1}{16\pi }(\xi-1/6)^2R^2(\tau),
\end{eqnarray}
that coincides with formula $(2.29)$ of \cite{v85}.

Using the formula $R=12H^2+6\frac{dH}{dt}$, one has approximately at late times $R\cong
-\frac{4M}{t}\sin(Mt)$. Then averaging over a period of oscillation one finally obtains
\begin{eqnarray}\label{a63}
\overline{\frac{1}{a^3}\frac{d}{dt}(a^3N(\tau;-\infty))}\cong
\frac{M^2}{2\pi t^2 }(\xi-1/6)^2.
\end{eqnarray}

At this moment an important remark is in order. If one makes, at
late times, the approximation  $a(t)\cong t^{2/3}$ (that corresponds
to the expansion law of a matter-dominated universe) one has $t\sim
\eta^3$, then $R(\eta)\sim \eta^{-3}$ what means that, at late
times, $V_{\bf k}(\eta)\sim \eta$  and thus one has an infinite
production of particles. One can avoid this problem neglecting the
effect of power-law expansion and  taking as a scalar factor the
function $a(t)\cong [1+\frac{2}{3Mt}\sin(Mt)]$, in that case one has
$t\sim \eta$ and thus, at late times, one has $V_{\bf k}(\eta)\sim
\eta^{-1}$.

To calculate the energy density per unit time one has to take into
account that the energy density at a given time, in general,
diverges for a no-conformally coupled fields (see for example
\cite{b80}), then one has to re-normalize this quantity. Moreover,
the re-normalized energy density depends, in general, on the
regularization method used, and in the final result it appears some
vacuum polarization terms, that are very difficult to separate of
those relative to the particle production (only in the conformally
coupled case this is possible to do).

For these reasons in order  to ``calculate'' the energy density per
unit we use the fact that the main contribution to the particle
production  come from modes with $k\sim M/2$. This can  be deduced
taking into account the oscillating behavior of the curvature at
late times, effectively, inserting $V_{\bf k}\cong
(\xi-1/6)\frac{4M}{\eta}\sin(M\eta)$ in the beta Bogoliubov
coefficient, one deduces that the oscillating character of the
integral disappear when $k= M/2$. Then one might concludes that an
approximation for the energy density per unit time, averaged over an
oscillation, is given by
\begin{eqnarray}\label{a64}
\overline{\frac{d}{dt}\rho(\tau;-\infty)}\cong
\frac{M}{2}\overline{\frac{1}{a^3}\frac{d}{dt}(a^3N(\tau;-\infty))}\cong
\frac{M^3}{4\pi t^2 }(\xi-1/6)^2.
\end{eqnarray}

At this point is important to remember that Starobinsky proposed in
\cite{s80} that the oscillating behavior of the scale factor can be
though as a coherent oscillations of a massive field described by
particles of mass $M$ (scalarons), then with this point of view,
gravitational particle production could be understood as a decay of
scalarons (due to its rapid oscillations) into other particles like
massless no-conformally coupled particles, massive  conformally
coupled ones, etc...

In this way, one can calculates the rate at which the energy of
scalarons is dissipated
\begin{eqnarray}\label{a65}
 \Gamma\equiv \frac{\overline{\frac{d}{dt}\rho(\tau;-\infty)}}{\overline{\rho(\tau)}},
\end{eqnarray}
where to obtain the energy density of the scalarons $\rho(\tau)$ one
must uses the ``effective'' Friedmann equation  $H^2=\frac{8\pi
\rho(\tau)}{3 m_p^2}$, being $m_p$ the Planck mass   (we say
``effective'' because  in the Starobinsky model the energy density
that appears in the equations is only due to the vacuum
polarization, but from Starobinsky's viewpoint, at late time, one
could thinks that scalarons drive the expansion of the universe).

Now since $\overline{H}=\frac{2}{3t}$ one  has
\begin{eqnarray}\label{a66}
 \Gamma\cong \frac{3}{2}\frac{M^3}{m_p^2}(\xi-1/6)^2.
\end{eqnarray}

Finally at time $t\sim \Gamma^{-1}$ the oscillations of the scalarons field are damped, the created particles thermalize, and the
universe becomes radiation dominated. To calculate its temperature, one has to use the
thermodynamical relation $\rho=\frac{\pi^2}{30}N(T) T^4$, where $N(T)$ is the effective number of relativistic degrees of freedom at temperature $T$,  the effective
Friedmann equation, and the fact that reheating ends when $\Gamma\sim H$ (see for example \cite{kls97}), then one obtains
\begin{eqnarray}\label{a67}
 T_{th}\sim \sqrt{\Gamma m_p}\sim |\xi-1/6|M^{3/2}m_p^{-1/2}.
\end{eqnarray}

\vspace{1cm}

 To end the section, we study massive conformally coupled particle production in the
 Starobinsky model. First at all note that, to apply  formula
 (\ref{a37}) one must assumes that $\lim_{t\rightarrow +\infty}a(t)=\lim_{t\rightarrow
 -\infty}a(t)$ \cite{bd79}, then one doesn't have to work with the Starobinsky model
 because it doesn't satisfy this assumption. We will work with the
 an scale factor that at early times is $a(t)\cong 1$ and at late
 time is $a(t)\cong [1+\frac{2}{3Mt}\sin(Mt)]$. We also assume that
 $m\ll M$, then since $\omega^2_{-,{\bf k}}=|{\bf k}|^2+m^2$, at late times, one has  $V_{\bf
 k}(\eta)=\omega^2_{-,{\bf k}}-\omega^2_{{\bf
 k}}(\eta)=m^2(1-a^2(\eta))\cong\frac{4m^2}{3M\eta}\sin(M\eta)$.

Now if one chooses the family of solutions
\begin{eqnarray}\label{a68}
 \chi_{\bf k, \tau}(\eta)=\frac{e^{-i\omega_{-,{\bf k}}\eta}}{\sqrt{2\omega_{-,{\bf k}}}}+\frac{1}{\omega_{-,{\bf k}}}
\int_{\tau}^{\eta}V_{\bf k}(\eta')\sin(\omega_{-,\bf
k}(\eta-\eta'))\chi_{\bf k,\tau}(\eta')d\eta',
\end{eqnarray}
one obtains
\begin{eqnarray}\label{a69}
 \beta_{\bf k}(\tau;-\infty)\cong -\frac{i}{2\omega_{-,{\bf k}}}\int_{\R} \theta(\tau-\eta)e^{-2i\omega_{-,{\bf k}}\eta} V_{\bf
k}(\eta)d\eta.
\end{eqnarray}

Thus, since now we cannot apply Plancherel's theorem,  to calculate
the density of produced particles at time $\tau$, we make the change
of variable $|{\bf k}|^2+m^2=x^2M^2$ and make the approximation
$xM\sqrt{x^2M^2-m^2}\cong x^2M^2-\frac{m^2}{2}$, then one gets
\begin{eqnarray}\label{a70}
N(\tau;-\infty)&&\cong
\frac{M^3}{2\pi^2a^3(\tau)}\left[\int_{m/M}^{\infty}x^2|\beta_{\bf
k}(\tau;-\infty)|^2dx-\frac{m^2}{M^2}\int_{m/M}^{\infty}|\beta_{\bf
k}(\tau;-\infty)|^2dx\right]\nonumber\\&& \cong
\frac{M^3}{2\pi^2a^3(\tau)}\int_{0}^{\infty}x^2|\beta_{\bf
k}(\tau;-\infty)|^2dx.
\end{eqnarray}

Now we can apply Plancherel's theorem to obtain
\begin{eqnarray}\label{a71}
N(\tau;-\infty)\cong\frac{1}{16\pi a^3(\tau)}\int_0^{\tau}V^2_{\bf
k}(\eta)d\eta,\end{eqnarray}
 and then, at late times, one has
\begin{eqnarray}\label{a72}
 \frac{1}{a^3}\frac{d}{dt}(a^3N(\tau;-\infty))\cong \frac{m^4}{9\pi M^2\tau^2}\sin^2(M\tau),
\end{eqnarray}
where we have used the approximation $a(\tau)\cong 1$. Averaging
over an oscillation one has
\begin{eqnarray}\label{a73}
\overline{\frac{1}{a^3}\frac{d}{dt}(a^3N(\tau;-\infty))}\cong
\frac{m^4}{18\pi M^2\tau^2}.
\end{eqnarray}

To calculate the energy density per unit time, we use once again
that the main contribution to the particle creation comes from modes
with $k\sim M/2$, then
\begin{eqnarray}\label{a74}
\overline{\frac{d}{dt}\rho(\tau;-\infty)}\cong
\frac{M}{2}\overline{\frac{1}{a^3}\frac{d}{dt}(a^3N(\tau;-\infty))}\cong
\frac{m^4}{36\pi M^2\tau^2}.
\end{eqnarray}

And finally, the rate at which the energy of scalarons is dissipated
is
\begin{eqnarray}\label{a75}
 \Gamma\equiv \frac{\overline{\frac{d}{dt}\rho(\tau;-\infty)}}{\overline{\rho(\tau)}}\cong
 \frac{m^4}{6 Mm_p^2},
\end{eqnarray}
this is  the result obtained by Starobinsky  \cite{s80}.

\vspace{1cm}

\subsection{Particle production by strong electromagnetic fields: Schwinger's formula}
Here we deduce the well-known Schwinger formula \cite{s51} for spin
and spinless particles
 using the W.K.B. approximation, that is, we compute the
probability that the vacuum state remains unchanged in the presence
of a constant electric field using the semi-classical approach.

First, we
consider, in the Minkowski space-time, the Klein-Gordon field in a box of volume $L^3$,
coupled with an external uniform vector potential
 ${\bf f}(t)$. The Klein-Gordon  equation
is equivalent to a Hamiltonian system, composed by an infinite
number of harmonic oscillators with frequencies which depend on
time. The mode equations are:
\begin{eqnarray}\label{a76} \chi_{\bf k}''+\omega_{\bf k}^2(t)\chi_{\bf k}=0\quad \mbox{with}\quad {\bf k}\in \Z^3,
\end{eqnarray}
where now
$
\omega_{\bf k}^2(t)=
\left|\frac{2\pi{\bf k}}{L}+{e}{\bf f}(t)\right|^2+m^2$ (being $e$ the electric charge).

We assume that $\lim_{t\rightarrow \pm\infty} \omega_{\bf k}(t)=
\omega_{\bf k,\pm}$,
and we
 write the ``in''-states as linear combinations
of the ``out''-states as follows
\begin{eqnarray}\label{a77}
\chi_{in,\bf k}(t)=\alpha_{\bf k}\chi_{out, \bf k}(t)+ \beta_{\bf
k}\chi^*_{out, \bf k}(t),\end{eqnarray} then one has,
$\hat{A}_{out,{\bf k}}=\alpha_{\bf k}\hat{A}_{in,{\bf k}}
+\beta^*_{\bf k}\hat{A}^{\dagger}_{in,{\bf k}}$.

Let  $|n_{\bf k}\rangle$ be the ``in''-state that contains
$n$ particles in the ${\bf k}$ mode,
 and let  $|n_{\bf k})$ be the ``out''-state that contains
 $n$ particles in the ${\bf k}$ mode.
Then, it is easy to obtain the following relations \cite{gmm94}:
\begin{eqnarray}\label{a78}
|0_{\bf k}\rangle=\tilde{C}_{\bf k}
\sum_{n=0}^{\infty}\left(\frac{\beta^*_{\bf k}}{\alpha^*_{\bf
k}}\right)^n|n_{\bf k}), \quad |0_{\bf k})= {C}_{\vec
k}\sum_{n=0}^{\infty} \left(-\frac{\beta^*_{\bf k}}{\alpha_{\bf
k}}\right)^n|n_{\bf k}
 \rangle,\end{eqnarray}
with $|\tilde{C}_{\bf k}|^2=|C_{\bf k}|^2=|\alpha_{\bf k}|^{-2}$.

From these relations, we deduce that the probability that a particle
in the ${\bf k}$ mode is produced  \cite{gmm94,p69,mp77}, is
\begin{eqnarray}\label{a79}
|(n_{\bf k}|0_{\bf k}\rangle|^2=\frac{|\beta_{\bf
k}|^{2n}}{(1+|\beta_{\bf k}|^2)^{n+1}},\end{eqnarray} and the
average number of produced particle in the ${\bf k}$ mode, is
\begin{eqnarray}\label{a80}
\langle 0_{\bf k}|\hat{a}_{out,{\bf k}}^+ \hat{a}_{out,{\bf
k}}|0_{\bf k}\rangle ={|\beta_{\bf k}|^2}.\end{eqnarray}

\begin{example}
 (Adiabatic approximation) From formulae (\ref{a29}) and
(\ref{a31}) one obtains the following formula for the number
density of produced particles
\begin{eqnarray}\label{d1}
{N}_{diag}(t;-\infty) =\frac{\alpha}{512\pi m}|{{\bf E}}(t)|,
\end{eqnarray}
where $\alpha=e^2$ is the fine structure constant and ${\bf
E}(t)=-\dot{\bf f}(t)$ is the external electric field. However,
their energy density diverges. To obtain a well-known defined
quantity one has to re-normalize the electric charge, after this
one has (see details in \cite{gmm94,h06,m79,cm89})
\begin{eqnarray}\label{d2}
{\rho}_{diag}(t;-\infty) = \frac{7\alpha^2} {1920\pi^2 m^4} |{{\bf
E}}(t)|^4 +\frac{\alpha} {960\pi^2m^2} \left(|\dot{{\bf
E}}(t)|^2-2{{\bf E}}(t)\cdot \ddot{{\bf E}}(t)\right).
\end{eqnarray}

Moreover it is also possible to calculate, after charge
re-normalization,  the induced electric field, that is, the
corrections to the external electric field due to vacuum
fluctuations \cite{gmm94, h06}
\begin{eqnarray}\label{d3}
{\bf E}_{vac}(t)= \frac{\alpha}{120\pi m^2} \ddot{{\bf E}}(t) -
\frac{7\alpha^2 }{360\pi m^5} {{\bf E}}(t) |{{\bf E}}(t)|^2,
\end{eqnarray}
and the effective Lagrangian density for the electric field
\cite{h06}
\begin{eqnarray}\label{d4}
{{\mathcal L}}_{eff}(t) =\frac{1}{8\pi}|{{\bf E}}(t)|^2+
\frac{7\alpha^2} {5760\pi^2 m^4} |{{\bf E}}(t)|^4 +\frac{\alpha}
{960\pi^2m^2} |\dot{{\bf E}}(t)|^2,
\end{eqnarray}
which generalizes the Euler-Heisenberg formula \cite{s51}, for a
time-dependent field.

\end{example}

From these results obtained above one can deduce that the probability that the
vacuum state remains unchanged, namely $P$ is
\begin{eqnarray}\label{a81}P=
\prod_{{\bf k}\in\Z^3} \frac{1}{1+|\beta_{\bf k}|^2}
=\exp\left(-\sum_{{\bf k}\in\Z^3} \log\left(1+|\beta_{\bf k}|^2
\right)\right) =\exp\left(-\sum_{{\bf k}\in\Z^3} \sum_{n=1}^{\infty}
\frac{(-1)^{n+1}}{n} |\beta_{\bf k}|^{2n} \right)\end{eqnarray}

As an application of this result, we'll find
 Schwinger's formula for scalar particles. Consider the case
${\bf f}(t)=(0,0,f(t))$, where
\begin{eqnarray}\label{a82}
f(t)=\left\{\begin{array}{ccc}
-E{T}&\mbox{ if } & t< -{T}\\
Et &\mbox{ if } & -{T}<t<{T}\\
E{T} &\mbox{ if } & t>{T},
\end{array}\right.\end{eqnarray}
being $E$ the electric field and with $T\gg 1$.
We suppose for example $eE>0$ (The case $eE<0$ is analogous).

The Schwinger formula gives the probability that the vacuum state
remains unchanged. Then, using the notation
\begin{eqnarray}\label{a83}
N\equiv \frac{2TL^3 E^2\alpha}{8\pi^3},\quad S\equiv \frac{\pi m^2}{
eE},\end{eqnarray}  Schwinger's formula for  spinless particles is
\cite{s51}
\begin{eqnarray}\label{a84}
P= \exp\left(-N\sum_{n=1}^{\infty}
\frac{(-1)^{n+1}}{n^2}\exp\left(-nS\right) \right).\end{eqnarray}

In order to deduce this formula,
we compute the $\beta$ Bogoliubov coefficient
using the
relativistic tunneling effect \cite{p72,hol99,ek88}, i.e., using the formulae given by the
 W.K.B. method in the complex plane \cite{f93,w73,m80}in a formal way. (we say in a formal way
because our field is not analytic, and the formula (\ref{a85}) that
we'll use is only mathematically justified for analytic fields). The
result is \cite{h03}
\begin{eqnarray} \label{a85}|\beta_{\bf k}|^2=\left\{\begin{array}{ccc}
\exp\left(- Im\int_{\gamma}\sqrt{\left(k_{\perp}^2+m^2+
\left(\frac{2\pi  k_3}{L}+eEz\right)^2\right)}dz
\right)& \mbox{if} & \left|\frac{2\pi  k_3}{L}\right|<  eE{T}\\
0& \mbox{if} & \left|\frac{2\pi  k_3}{L}\right|> eE{T},
\end{array}\right.\end{eqnarray}
where $k_{\perp}\equiv\frac{2\pi }{L}(k_1,k_2)$, and $\gamma$ is a
simple curve in the complex plane, containing the complex turning
points $\frac{-\frac{2\pi  k_3}{L}\pm \sqrt{k^2_{\perp}+m^2}}{eE}$
as interior points (Note that for $\left|\frac{2\pi  k_3}{L}\right|>
eE{T}$ there isn't turning points \cite{h03a}, that's the reason why
for that modes its beta-Bogoliubov coefficient vanishes). Now, it's
easy to verify that
\begin{eqnarray} \label{a86}|\beta_{\bf k}|^2=\left\{\begin{array}{ccc}
\exp\left(-\frac{\pi(k_{\perp}^2+m^2)}{e E}
\right)& \mbox{if} & \left|\frac{2\pi  k_3}{L}\right|<  e E{T}\\
0& \mbox{if} & \left|\frac{2\pi  k_3}{L}\right|> e E{T},
\end{array}\right.\end{eqnarray}
and therefore, the probability that the vacuum state remains
unchanged is
\begin{eqnarray}\label{a87}&&P
=\exp\left(-\sum_{{\bf k}\in\Z^3} \sum_{n=1}^{\infty}
\frac{(-1)^{n+1}}{n} |\beta_{\bf k}|^{2n} \right)=
\exp\left(-N\sum_{n=1}^{\infty}
\frac{(-1)^{n+1}}{n^2}\exp\left(-nS\right) \right),\end{eqnarray} in
agreement with  Schwinger's result.

In the same way one obtains that the
average number of produced pairs per unit  volume and per unit  time, is
\begin{eqnarray}\label{a88}\frac{E^2\alpha}{8\pi^3}\exp\left(
-\frac{\pi m^2}{ e E} \right).
\end{eqnarray}

To obtain Schwinger's formula  for fermions, one has to use the
Pauli Exclusion Principle to get the following relation between
Bogoliubov's coefficients \cite{n70,n70a,h03}
\begin{eqnarray}\label{a89}
|\alpha_{\bf k}|^2+|\beta_{\bf k}|^2=1.\end{eqnarray} 

 Now
$|\beta_{\bf k}|^2$ is the probability that a particle is created in
the ${\bf k}$ mode, because it's the same that the average number of
produced particles in that mode, and consequently $|\alpha_{\bf
k}|^2$ is the probability that no particles are produced in that
mode. Therefore, due to the existence of two states with spin $1/2$,
the probability that the vacuum state remains unchanged is
\begin{eqnarray}\label{90}P=
\prod_{{\bf k}\in\Z^3} (1-|\beta_{\bf k}|^2)^2
=\exp\left(-2\sum_{{\bf  k}\in\Z^3}
\sum_{n=1}^{\infty} \frac{1}{n} |\beta_{\bf
k}|^{2n}\right)=\exp\left(-2N\sum_{n=1}^{\infty}
\frac{1}{n^2}\exp\left(-nS\right) \right).\end{eqnarray}

\begin{remark}
 In that Section we've obtained Schwinger's formula in an easy but formal way. However if  one wants a complete
demonstration of the formula with all the details, one can look up
at \cite{gg95,m07,h04,bgs75}. Essentially,
the idea is to make the change of variable
 $y=\sqrt{\frac{2}{e E}}(\frac{2\pi k_3}{L}+eEt)$, then the Klein-Gordon
equation behaves
\begin{eqnarray}\label{c1}
{\chi}_{\bf k}''+\tilde{\omega}_{\bf k}^2(y)\chi_{\bf k}=0,
\end{eqnarray}
where, $\tilde{\omega}_{\bf k}(y)\equiv
\sqrt{\frac{1}{4}y^2-A_{\bf k}}$ and $A_{\bf
k}\equiv\frac{-1}{2eE}(k_{\perp}^2 +m^2)$.

In this case, the function $\chi_{{\bf
k},0;WKB}(y;\epsilon=1)\equiv
 \sqrt{\frac{1}{2\tilde{\omega}_{\bf k}(y)}}e^{-i\int^{y}\tilde{\omega}_{\bf k}(\tau)d\tau},
$ has the asymptotic behavior
\begin{eqnarray}\label{c2}
\chi_{{\bf k},0;WKB}(y;\epsilon=1)=\left\{\begin{array}{cc}
e^{iy^2/4}(y^2)^{-1/4-iA_{\bf k}/2}& y\rightarrow -\infty\\
   e^{-iy^2/4}(y^2)^{-1/4+iA_{\bf k}/2}& y\rightarrow \infty.                              \end{array}
\right.
\end{eqnarray}

Note that $\chi_{{\bf
k},0;WKB}'(y;\epsilon=1)=-i\tilde{\omega}_{\bf k}(y)\chi_{{\bf
k},0;WKB}(y;\epsilon=1)$ when $y\rightarrow \pm\infty$, this means
that the diagonalization metod and zero order adiabatic vacuum
modes give the same value for $\beta_{\bf k}$.

On the other hand, a independent set of solutions of (\ref{c1}) is
given by the two following  parabolic cylinder functions
\cite{as72}
\begin{eqnarray}\label{c3}
u_{{\bf k},1}(y)=\exp\left(-\frac{i}{4}y^2\right)M\left(
-\frac{i}{2}A_{\bf
k}+\frac{1}{4},\frac{1}{2},\frac{i}{2}y^2\right)
\end{eqnarray}
\begin{eqnarray}\label{c4}
u_{{\bf
k},2}(y)=\frac{1}{\sqrt{2}}\exp\left(-\frac{i}{4}y^2\right) y
\exp\left(-\frac{i\pi}{4}\right)M\left( -\frac{i}{2}A_{\bf
k}+\frac{3}{4},\frac{3}{2},\frac{i}{2}y^2\right),\end{eqnarray}
where $M$ is the Kummer's function.

We now define the mode solution
\begin{eqnarray}\label{c5}
\chi_{\bf k}(y)\equiv \hbar^{1/2}B_{\bf k}^{-1}e^{i\pi/8}e^{-\pi
A_{\bf k}/4}2^{-1/4-iA_{\bf k}/2}\varphi_{\bf k}(y),
\end{eqnarray}
with
\begin{eqnarray}\label{c6}
\varphi_{\bf k}(y)\equiv\frac
{\Gamma\left(\frac{1}{4}+\frac{i}{2}A_{\bf
k}\right)}{\Gamma\left(\frac{1}{2}\right)}u_{{\bf k},1}(y)+\frac
{\Gamma\left(\frac{3}{4}+\frac{i}{2}A_{\bf
k}\right)}{\Gamma\left(\frac{3}{2}\right)}u_{{\bf k},2}(y),
\end{eqnarray}
and
\begin{eqnarray}\label{c7}
B_{\bf k}\equiv\frac{\Gamma\left(\frac{1}{4}+\frac{i}{2}A_{\bf
k}\right)} {\Gamma\left(\frac{1}{4}-\frac{i}{2}A_{\bf k}\right)}+i
\frac{\Gamma\left(\frac{3}{4}+\frac{i}{2}A_{\bf k}\right)}
{\Gamma\left(\frac{3}{4}-\frac{i}{2}A_{\bf k}\right)}.
\end{eqnarray}

Then, from the asymptotic behavior of the Kummer's function
\cite{as72} we can see that
\begin{eqnarray}\label{c8}
\chi_{\bf k}(y)\rightarrow \chi_{{\bf k},0;WKW}(y;\epsilon=1),
\quad \mbox{when}\quad y\rightarrow -\infty,
\end{eqnarray}
and
\begin{eqnarray}\label{c9}
\chi_{\bf k}(y)\rightarrow B_{\bf k}^{-1}e^{i\pi/4}2^{1-iA_{\bf
k}}\chi_{{\bf k},0;WKW}(y;\epsilon=1)+ C_{\bf k}B_{\bf
k}^{-1}\chi_{{\bf k},0;WKW}^*(y;\epsilon=1), \quad
\mbox{when}\quad y\rightarrow \infty,
\end{eqnarray}
with
$$C_{\bf k}\equiv\frac{\Gamma\left(\frac{1}{4}+\frac{i}{2}A_{\bf k}\right)}
{\Gamma\left(\frac{1}{4}-\frac{i}{2}A_{\bf k}\right)}-i
\frac{\Gamma\left(\frac{3}{4}+\frac{i}{2}A_{\bf k}\right)}
{\Gamma\left(\frac{3}{4}-\frac{i}{2}A_{\bf k}\right)}.
$$

Then from this last formula we can deduce that, in both cases
(diagonalization method and zero order adiabatic vacuum modes),
 the square of the $\beta$-Bogoliubov coefficient is given  by

\begin{eqnarray}\label{c10}
|\beta_{\bf k}|^2={|C_{\bf k}/B_{\bf k}|^2}=e^{2\pi A_{\bf k}}
=\exp\left(-\frac{\pi}{eE}(k_{\perp}^2 +m^2)\right),\end{eqnarray}
in agreement with (\ref{a86}).
\end{remark}

\subsection{Moving Mirrors}
Consider a massless scalar field $\phi$  in the $2$-dimensional
Minkowski space-time interacting with a perfect reflecting moving
mirror. Assume that the mirror trajectory follows an inertial
prescribed trajectory $x=g(t)$, that in light-like coordinates
$u\equiv t-z$ and $v\equiv t+z$, we write as $v=V(u)$ or $u=U(v)$.

For a perfectly reflecting mirror the set of ``in'' and ``out'' mode
functions is \cite{dw75,fd76,c02b}
\begin{eqnarray}\label{a91}\left\{\begin{array}{c}
\phi_{\omega,R}^{in}(u,v)=\frac{1}{\sqrt{4\pi|\omega| }}
\left(e^{-i\omega v}-e^{-i\omega V(u)}\right)\theta(v-V(u))\\
\\
\phi_{\omega,L}^{in}(u,v)=\frac{1}{\sqrt{4\pi|\omega| }}
\left(e^{-i\omega u}-e^{-i\omega U(v)}\right)\theta(u-U(v))
\end{array}\right.
\end{eqnarray}
\begin{eqnarray}\label{a92}\left\{\begin{array}{c}
\phi_{\omega,R}^{out}(u,v)=\frac{1}{\sqrt{4\pi|\omega| }}
\left(e^{-i\omega u}-e^{-i\omega U(v)}\right)\theta(v-V(u))\\
\\
\phi_{\omega,L}^{out}(u,v)=\frac{1}{\sqrt{4\pi|\omega| }}
\left(e^{-i\omega v}-e^{-i\omega V(u)}\right)\theta(u-U(v)),
\end{array}\right.
\end{eqnarray}
where $\phi_{\omega,R}^{in}$ (resp. $\phi_{\omega,L}^{in}$)
represents particles with frequency $\omega$ coming from the right
(resp. left) null past infinity domain ${\mathcal J}_R^-$ (resp.
${\mathcal J}_L^-$), and $\phi_{\omega,R}^{out}$ (resp.
$\phi_{\omega,L}^{out}$) represents particles with frequency
$\omega$ going to the right (resp. left) null future infinity domain
${\mathcal J}_R^+$ (resp. ${\mathcal J}_L^+$).

It is a well-known fact that the average number of particles in the
$\omega$ mode produced from the vacuum in the right hand side (rhs)
of the mirror, is \cite{c02b,w85}
\begin{eqnarray}\label{a93}
N(\omega)=\int d\omega' |\beta^{R,R}_{\omega,\omega'}|^2.
\end{eqnarray} 

Then
our main objective  is to calculate the beta Bogoliubov coefficient
\begin{eqnarray}\label{a94}
\beta^{R,R}_{\w,\w'}\equiv
{({\phi_{\omega,R}^{out}}^*;\phi_{\omega',R}^{in})}^*, \quad \mbox{
with } \quad \w,\w'>0,
\end{eqnarray}
where the parenthesis in the right member denotes the usual product
for scalar fields \cite{bd82}, that is,
\begin{eqnarray}\label{a95}
\beta^{R,R}_{\omega,\omega'} \equiv i\int_{g(t)}^{\infty}
{\phi}^{out}_{\omega, R}(t,x)\overleftrightarrow{\partial_t}
{\phi}^{in}_{\omega',R}(t,x) d{x},
\end{eqnarray}
that doesn't depend on the space-like hyper-surface chosen.

The best way to perform this calculation is to choose as
hyper-surface the right null future infinity ${\mathcal J}_R^+$,
then one has
\begin{eqnarray}\label{a96}
\beta^{R,R}_{\w,\w'}=2i\int_{\R}du\phi_{\w,R}^{out}\partial_u\phi_{\w',R}^{in}=
\frac{-i}{{2\pi\sqrt{\omega\omega'} }}\int_{\R}due^{-i\omega
u}\partial_u e^{-i\omega'
V(u)}=\frac{1}{2\pi}\sqrt{\frac{\w}{\w'}}
\int_{\R}due^{-i\w u}e^{-i\w' V(u)}.
\end{eqnarray}

 Assuming that the mirror's velocity converges fast enough to a
constant when $|u|\rightarrow\infty$, and integrating by parts one
gets
\begin{eqnarray}\label{a97}
\beta_{\w,\w'}^{R,R}=-\frac{1}{2\pi i}\sqrt{\w\w'}
\int_{\R}du\frac{V''(u)}{(\w+\w' V'(u))^2} e^{-i\w u}e^{-i\w' V(u)}.
\end{eqnarray}

For simplicity  we assume that the mirror's acceleration is
discontinuous at the point $u=a$, after another integration by parts
one obtains
\begin{eqnarray}\label{a98}
\beta_{\w,\w'}^{R,R}&&=-\frac{1}{2\pi }\sqrt{\w\w'} \frac{1}{(\w+\w'
V'(a))^3} e^{-i\w a}e^{-i\w' V(a)}(V''(a^-)-V''(a^+)) \nonumber\\&&
+\frac{1}{2\pi }\sqrt{\w\w'} \int_{\R}du\left[\frac{V'''(u)}{(\w+\w'
V'(u))^3}- \frac{3\w' (V''(u))^2}{(\w+\w' V'(u))^4}\right] e^{-i\w
u}e^{-i\w' V(u)}.
\end{eqnarray}

From this formula,  assuming that the mirror's trajectory is
asymptotically inertial, i.e., $V'(u)>0 \quad \forall u\in \R$ (see
for example \cite{w85}), one concludes that
$\left|\beta_{\w,\w'}^{R,R}\right|^2$ and $
\w\left|\beta_{\w,\w'}^{R,R}\right|^2$ are integrable functions in
the domain $[0,\infty)^2\setminus [0,1]^2$.

Finally, we're interested in the production of particles in the
infrared domain, i.e., we want to calculate
$\left|\beta_{\w,\w'}^{R,R}\right|^2$ in $[0,1]^2$.
We write the Bogoliubov coefficient as follows
\begin{eqnarray}\label{a99}
\beta_{\w,\w'}^{R,R}=\frac{1}{2\pi}\sqrt{\frac{\w}{\w'}}
\int_{\R}due^{-i(\w+B\w') u}e^{-i\w' (V(u)-Bu)},
\end{eqnarray}
with $B>0$. After an integration by parts one obtains
\begin{eqnarray}\label{a100}
\beta_{\w,\w'}^{R,R}=-\frac{1}{2\pi}\frac{\sqrt{\w\w'}}{\w+B\w'}
\int_{\R}du(V'(u)-B)e^{-i\w u}e^{-i\w' V(u)},
\end{eqnarray}
and thus, if the function $|V'(u)-B|$ is integrable in $\R$ for some
$B>0$, it can be deduced that  $\left|\beta_{\w,\w'}^{R,R}\right|^2$
and $\w\left|\beta_{\w,\w'}^{R,R}\right|^2$ are integrable
functions in the domain $[0,1]^2$. An example of this kind of
trajectories is
\begin{eqnarray}\label{a101}
V(u)=\left\{\begin{array}{cc} Bu & u\leq 0 \\
V(u) & 0\leq u\leq u_0\\
V(u_0)+B(u-u_0) & u\geq u_0. \end{array}\right.
\end{eqnarray}

However if we one is only interested  in the convergence of the
function $\w\left|\beta_{\w,\w'}^{R,R}\right|^2$ in the domain
$[0,1]^2$, one only needs trajectories that satisfy
\begin{eqnarray}\label{a102}
\int_{-\infty}^0 du|V'(u)-B_1|<\infty\mbox{ and
}\int^{\infty}_0du|V'(u)-B_2|<\infty
\end{eqnarray}
 for some no-negatives constants $B_1$ and
$B_2$, (here it's important to remark that one of  these constants
can be zero, that is, it's not worth that the trajectory be
asymptotically inertial). To prove this statement we write
\begin{eqnarray}\label{a103}
\beta_{\w,\w'}^{R,R}=\frac{1}{2\pi}\sqrt{\frac{\w}{\w'}}\left[
\int_{-\infty}^0due^{-i(\w+B_1\w') u}e^{-i\w'
(V(u)-B_1u)}+\int_0^{\infty}due^{-i(\w+B_2\w') u}e^{-i\w'
(V(u)-B_2u)}\right],
\end{eqnarray}
and we assume for simplicity that $V(0)=0$. After an integration by
parts one gets the formula
\begin{eqnarray}\label{a104}
\beta_{\w,\w'}^{R,R}&=&-\frac{1}{2\pi}\sqrt{\w\w'}
\frac{B_1-B_2}{(\w+B_1\w')(\w+B_2\w')}-\frac{1}{2\pi}\frac{\sqrt{\w\w'}}{\w+B_1\w'}
\int_{-\infty}^0du(V'(u)-B_1)e^{-i\w u}e^{-i\w'
V(u)}\nonumber\\&&-\frac{1}{2\pi}\frac{\sqrt{\w\w'}}{\w+B_2\w'}
\int_{0}^{\infty}du(V'(u)-B_2)e^{-i\w u}e^{-i\w' V(u)},
\end{eqnarray}
that proves the statement.

In conclusion we've proved that for asymptotically inertial
trajectories with continuous velocity the radiated energy is finite.
However it is also possible an infinite production  of particles
with very low frequency (an infrared divergency). To remove this
divergency one must assumes that the initial and final mirror's
velocity is the same.

Note that particle creation for  partially transmitting mirrors is a
bit different: At very high frequencies the mirror behaves
transparent, and then there are not particle production
independently of the mirror's trajectory. On the other hand, at very
low frequencies the mirror behaves a perfect reflector and then the
same kind of infrared problems as in the perfect reflector case
remain. Consequently, if one is only interested in a finite radiated
energy, one can considers trajectories with a continuous velocity
$V'(u)\quad \forall u\in\R$, which fulfill the condition (\ref{a102})
as, for instance,   the no-asymptotically inertial trajectory:
\begin{eqnarray}\label{a105}
v=V(u)\equiv\left\{\begin{array}{ccc}
u&\mbox{if}& u\leq 0\\
&&\\
\frac{1}{k}(1-e^{-ku})&\mbox{if}& u\geq 0.
\end{array}\right.\end{eqnarray}

\subsubsection{Simulating black body collapse}

Now we're interested in  a trajectory that
 simulates a black body collapse
\cite{fd77,bd82,cw87}, that is, with the following form
\begin{eqnarray}\label{106}
v=V(u)\equiv\left\{\begin{array}{ccc}
u&\mbox{if}& u\leq 0\\
&&\\
\frac{1}{k}(1-e^{-ku})&\mbox{if}& 0\leq u\leq u_0\\
&&\\
V(u_0)+A(u-u_0)&\mbox{if}& u\geq u_0,
\end{array}\right.\end{eqnarray}
with $A=e^{-ku_0}$, where $k$ is a frequency and $ku_0\gg 1$.

Note that this trajectory can  be written under the following form,
too
\begin{eqnarray}\label{a107}
u=U(v)\equiv\left\{\begin{array}{ccc}
v&\mbox{if}& v\leq 0\\
&&\\
-\frac{1}{k}\ln(1-k v)&\mbox{if}& 0\leq v\leq v_0\\
&&\\
U(v_0)+A^{-1}(v-v_0)&\mbox{if}& v\geq v_0.
\end{array}\right.\end{eqnarray}

Then,
\begin{eqnarray}\label{a108}
\beta^{R,R}_{\w,\w'}=2i\int_{\R}du\phi_{\w,R}^{out}\partial_u\phi_{\w',R}^{in}&=&
\frac{1}{2\pi i\sqrt{\w\w'}}\frac{\w'}{\w+\w'}
     -\frac{1}{2\pi
i\sqrt{\w\w'}}e^{-i\w
u_0}e^{-i\w'V(u_0)}\frac{\w'A}{\w+\w'A}\nonumber\\&& -\frac{1}{2\pi}
\sqrt{\frac{\w'}{\w}}\frac{1}{k}
\int_{0}^{1-A}ds(1-s)^{i{\w}/{k}}e^{-i{\w'}/{k}s}.
\end{eqnarray}

Assuming for simplicity $\w\sim k$, 
if one
 is interested in the domain of frequencies  $1\ll {\w'}/{k}\ll A^{-1}$, one arrives at
\begin{eqnarray}\label{a109}
\beta^{R,R}_{\w,\w'}&\cong & \frac{1}{2\pi i\sqrt{\w\w'}}
      -\frac{1}{2\pi} \sqrt{\frac{\w'}{\w}}\frac{1}{k}
\int_{0}^{1-A}ds(1-s)^{i{\w}/{k}}e^{-i{\w'}/{k}s}.
\end{eqnarray}

To obtain an explicit expression to the second term on the rhs, we
consider the domain
\begin{eqnarray}\label{a110}
D\equiv \{z\in \C/ \Re z\in [0,1-A], \Im z \in [-\epsilon,0], \mbox{
with } \frac{k}{\w'}\ll \epsilon\ll 1 \}.\end{eqnarray} and, going
through the same steps as in \cite{ha05}, one easily obtains
\begin{eqnarray}\label{a111}
\beta^{R,R}_{\w,\w'}\cong \frac{1}{2\pi i
\sqrt{\w\w'}}e^{-i{\w'}/{k}}\left(\frac{ik}{\w'}\right)^{i{\w}/{k}}
\Gamma\left(1+i{\w}/{k}\right),
\end{eqnarray}
and thus, using that
$|\Gamma\left(1+i{\w}/{k}\right)|^2=\frac{\pi{\w}/{k}}{\sinh\left(\pi{\w}/{k}
 \right)}$ (see \cite{as72}) one gets, for a
 perfect reflecting mirror, that
\begin{eqnarray}\label{a112}
\left|\beta^{R,R}_{\w,\w'}\right|^2\cong \frac{1}{2\pi \w'
k}\left(e^{2\pi{\w}/{k}}-1\right)^{-1},
\end{eqnarray}
in the range $1\ll {\w'}/{k}\ll A^{-1}$.

\begin{remark} From that formula, one deduces that the number of radiated
particles in the $\w$ mode diverges logarithmically with
$u_0\rightarrow \infty$. In this situation the physically relevant
quantity is the number of created particles in the $\w$ mode per
unit time. This dimensionless quantity is finite and its value is
given by \cite{n03,ha05}
\begin{eqnarray}\label{a113}
\lim_{u_0\rightarrow \infty}\frac{1}{u_0}{
N}(\w)=\frac{1}{2\pi}\left( e^{2\pi\w/k}-1\right)^{-1}.
\end{eqnarray}
\end{remark}

Now we'll study what happens when the mirror is partially reflecting.
First, we search for the co-moving coordinates $(\tau, \rho)$, that
is, the coordinates for which the mirror is at rest, $\tau$ being i
the proper time of the mirror, and we take $\rho$ such that its
trajectory is given by $\rho=0$. Introducing the light-like
coordinates  $(\bar{u},\bar{v})$ defined as
\begin{eqnarray}\label{a114}
\bar{u}\equiv \tau-\rho;\quad \bar{v}\equiv \tau+\rho,
\end{eqnarray}
we will calculate the mirror's trajectory in the coordinates
$(\bar{u},\bar{v})$.  Along this trajectory, the length element
obeys the identity \cite{op01}
\begin{eqnarray}\label{a115}
d\tau^2=d\bar{u}^2=d\bar{v}^2=V'(u)du^2=U'(v)dv^2.
\end{eqnarray}

Then, an easy calculation yields the relations
\begin{eqnarray}\label{a116}
\bar{v}=\bar{u}(u)\equiv\left\{\begin{array}{ccc} u&\mbox{if}& u\leq 0\\
&&\\
\frac{2}{k}(1-e^{-k{u}/{2}})&\mbox{if}& 0\leq u\leq u_0\\
&&\\
\bar{u}(u_0)+\sqrt{A}(u-u_0)&\mbox{if}& u\geq u_0,
\end{array}\right.\end{eqnarray}
and,
\begin{eqnarray}\label{a117}
\bar{u}=\bar{v}(v)\equiv\left\{\begin{array}{ccc}v&\mbox{if}& v\leq 0\\
&&\\
\frac{2}{k}(1-\sqrt{1-kv})&\mbox{if}& 0\leq v\leq v_0\\
&&\\
\bar{v}(v_0)+A^{-{1}/{2}}(v-v_0)&\mbox{if}& v\geq v_0.
\end{array}\right.\end{eqnarray}

When the mirror is at rest,  scattering is described by  the
S-matrix (see \cite{he06a,jr91,jr92} for more details)
\begin{eqnarray}\label{a118}S(\w)=\left(\begin{array}{cc}
{s}(\w)&{r}(\w)e^{-2i\w L}\\
{r}(\w)e^{2i\w L}&{s}(\w)\end{array}\right),\end{eqnarray} where
$x=L$ is the position of the mirror. The $S$ matrix is taken to be
real in the temporal domain, causal, unitary, and the identity at
high frequencies \cite{jr91}. Correspondingly,  the "in" modes in
the coordinates $(\bar{u},\bar{v})$ are \cite{bc95}
\begin{eqnarray}\label{a119}
g^{in}_{\w, R}(\bar{u},\bar{v})=
\frac{1}{\sqrt{4\pi|\omega| }}s(\w)e^{-i\omega \bar{v}}\theta(\bar{u}-\bar{v})+\frac{1}{\sqrt{4\pi|\omega| }}
\left(e^{-i\omega \bar{v}}+r(\w)e^{-i\omega \bar{u}}\right)\theta(\bar{v}-\bar{u}).
\end{eqnarray}

\begin{eqnarray}\label{a120}
g^{in}_{\w, L}(\bar{u},\bar{v})=\frac{1}{\sqrt{4\pi|\omega| }}
\left(e^{-i\omega \bar{u}}+r(\w)e^{-i\omega \bar{v}}\right)\theta(\bar{u}-\bar{v})+
\frac{1}{\sqrt{4\pi|\omega| }}s(\w)e^{-i\omega \bar{u}}\theta(\bar{v}-\bar{u}).
\end{eqnarray}

On the other hand, the "in" modes in the coordinates $(u,v)$, namely
$\phi^{in}$, are defined in the right null past infinity ${\mathcal
J}^-_R$ by
\begin{eqnarray}\label{a121}
\phi^{in}_{\w,R}=\frac{1}{\sqrt{4\pi|\omega| }}e^{-i\w v}; \quad
\phi^{in}_{\w,L}=0,
\end{eqnarray}
and in the left null past infinity ${\mathcal J}^-_L$ by
\begin{eqnarray}\label{a122}
\phi^{in}_{\w,R}=0; \quad
\phi^{in}_{\w,L}=\frac{1}{\sqrt{4\pi|\omega| }}e^{-i\w u}.
\end{eqnarray}

Writing $\bar{g}^{in}_{\w, k}(u,v)\equiv g^{in}_{\w,
k}(\bar{u}(u),\bar{v}(v))$ with $k=R,L$, and using  that
$\bar{g}^{in}_{-\w, k}=\bar{g}^{in
*}_{\w, k}$, one obtains the following relation
\begin{eqnarray}\label{a123}
\phi^{in}_{\w,k}=\int_{\R}d\w'\chi(\w')(\bar{g}^{in}_{\w',
k};\phi^{in}_{\w,k})\bar{g}^{in}_{\w', k},\quad \mbox{ with } \quad
k=R,L
\end{eqnarray}
with $\chi(\w')$  the sing function. To calculate explicitly the
"in" modes, I choose the coefficients
 $r(\w)=\frac{-i\gamma}{\w+i\gamma}$ and $s(\w)=\frac{\w}{\w+i\gamma}$
with $\alpha\geq 0$ that correspond to the so-called
Barton-Calogeracos model \cite{bc95,n01, c02a}. In this case, on the
rhs of mirror one has \cite{he08a}
\begin{eqnarray}\label{a124} 
\phi^{in}_{\w,R}(u,v)=\frac{1}{\sqrt{4\pi|\omega| }}e^{-i\omega v}+
\phi_{\w, R}^{refl}(u);\qquad \phi^{in}_{\w,L}(u,v)=\phi_{\w,
L}^{trans}(u),
\end{eqnarray}
where
\begin{eqnarray}\label{a125}\hspace{-1cm}
\phi_{\w, R}^{refl}(u)=\left\{
\begin{array}{cc}
\frac{1}{\sqrt{4\pi|\omega| }}\frac{-i\gamma}{\w+i\gamma}e^{-i\w V(u)};&u\leq 0\\
&\\
 \frac{1}{\sqrt{4\pi|\omega| }}\frac{-i\gamma}{\w+i\gamma}e^{-\gamma\bar{u}(u)}
 -\frac{2\gamma}{k\sqrt{4\pi|\omega|
}}e^{-i\frac{\w}{k}} \int_0^{\frac{k}{2}\bar{u}(u)}dse^{\frac{i\w
}{k}\left(s+1-\frac{k}{2}\bar{u}(u)\right)^2}
e^{-\frac{2\gamma s}{k}};& 0\leq u\leq u_0\\
&\\
\frac{1}{\sqrt{4\pi|\omega| }}\frac{-i\gamma}{\w+i\gamma}e^{-\gamma\bar{u}(u)}
 -\frac{1}{\sqrt{4\pi|\omega|
}}\frac{i\gamma}{\sqrt{A}\w+i\gamma}\left[ e^{-i\w V(u)}-e^{-i\w
V(u_0)}e^{-\gamma(\bar{u}(u)-\bar{u}(u_0))}
\right]& \\
& \\ -\frac{2\gamma}{k\sqrt{4\pi|\omega|
}}e^{-i\frac{\w}{k}}e^{-\gamma(\bar{u}(u)-\bar{u}(u_0))}
\int_0^{\frac{k}{2}\bar{u}(u_0)}dse^{\frac{i\w
}{4}\left(s+1-\frac{k}{2}\bar{u}(u_0)\right)^2} e^{-\frac{2\gamma
s}{k}};& u\geq u_0
\end{array}\right.
\end{eqnarray}
and
\begin{eqnarray}\label{a126}\hspace{-1cm}
\phi_{\w, L}^{trans}(u)=\left\{
\begin{array}{cc}
\frac{1}{\sqrt{4\pi|\omega| }}\frac{\w}{\w+i\gamma}e^{-i\w V(u)}
;&u\leq 0\\
&\\
\frac{1}{\sqrt{4\pi|\omega| }}e^{-i\w u}
 +\frac{1}{\sqrt{4\pi|\omega| }}\frac{-i\gamma}{\w+i\gamma}e^{-\alpha\bar{u}(u)}
 -\frac{2\gamma}{k\sqrt{4\pi|\omega|
}} \int^{\frac{k}{2}\bar{u}(u)}_0ds (s+1-\frac{k}{2}\bar{u}(u)
)^{2i\frac{\w}{k}} e^{-\frac{2\gamma s
}{k}};& 0\leq u\leq u_0\\
&\\\frac{1}{\sqrt{4\pi|\omega| }}\frac{-i\gamma}{\w+i\gamma} e^{-\gamma\bar{u}(u)}
+\frac{1}{\sqrt{4\pi|\omega|}}\frac{e^{-i\w u_0}}{\w+i\gamma\sqrt{A}}\left[ \w
e^{-i\frac{\w}{\sqrt{A}}(\bar{u}(u)-\bar{u}(u_0))}+i\gamma\sqrt{A}
e^{-\gamma(\bar{u}(u)-\bar{u}(u_0))}\right]& \\
& \\
-\frac{2\gamma}{k\sqrt{4\pi|\omega|
}}e^{-\gamma(\bar{u}(u)-\bar{u}(u_0))}
\int^{\frac{k}{2}\bar{u}(u_0)}_0ds (s+1-\frac{k}{2}\bar{u}(u_0)
)^{2i\frac{\w}{k}} e^{-\frac{2\gamma s }{k}} ;& u\geq u_0
\end{array}\right.
\end{eqnarray}

Note that  in the case of perfect reflection, that is, when
$\gamma\rightarrow \infty$ one has
\begin{eqnarray}\label{a127}
\phi_{\w, R}^{refl}(u)\rightarrow -\frac{1}{\sqrt{4\pi|\omega| }}e^{-i\w V(u)};\qquad
\phi_{\w, L}^{trans}(u)\rightarrow 0,
\end{eqnarray}
and when the mirror is transparent, i.e., when $\gamma\rightarrow 0$
one has
\begin{eqnarray}\label{a128}
\phi_{\w, R}^{refl}(u)\rightarrow 0;\qquad
\phi_{\w, L}^{trans}(u)\rightarrow \frac{1}{\sqrt{4\pi|\omega| }}e^{-i\w u}.
\end{eqnarray}

Since we're interested in the particle production on the rhs of the
mirror, we must now calculate
\begin{eqnarray}\label{a129}
 \beta^{R,R}_{\w,\w'}\equiv {({\phi_{\omega,R}^{out}}^*;\phi_{\omega',R}^{in})}^*,\quad \mbox{ and
 }\quad
\beta^{R,L}_{\w,\w'}\equiv {({\phi_{\omega,R}^{out}}^*;\phi_{\omega',L}^{in})}^*
\quad \w,\w'>0.
\end{eqnarray}

In order to calculate this products we choose the right null infinity
${\mathcal J}^+_R$, because here the "out" modes have a very easy
form, then
\begin{eqnarray}\label{a130}
 \beta^{R,R}_{\w,\w'}= {({\phi_{\omega,R}^{out}}^*;\phi_{\omega',R}^{refl})}^*,\quad \mbox{ and
 }\quad
\beta^{R,L}_{\w,\w'}\equiv {({\phi_{\omega,R}^{out}}^*;\phi_{\omega',L}^{trans})}^*.
\end{eqnarray}

We start calculating
$\beta^{R,R}_{\w,\w'}=2i\int_{\R}du\phi_{\omega,R}^{out}
\partial_u \phi_{\omega',R}^{refl}$, with the result
\begin{eqnarray}\label{a131}
\beta^{R,R}_{\w,\w'}&\cong&
\frac{1}{2\pi\sqrt{\w\w'}}\frac{\gamma}{\w'+i\gamma}\left[1
-\frac{\gamma}{k} \int_A^1dx
x^{i{\w}/{k}-{1}/{2}}e^{-{2\gamma}(1-\sqrt{x})/k}\right]\nonumber\\&&\hspace{-1cm}
+\frac{\gamma}{2\pi k i\sqrt{\w\w'}}e^{-i{\w'}/{k}} \int_A^1dx
x^{i{\w}/{k}-{1}/{2}}e^{i{\w'x}/{k}}\left[1
-\frac{2\gamma}{k}
\int_0^{1-\sqrt{x}}e^{i{\w'}(s^2+2s\sqrt{x})/k}e^{-{2\gamma
s}/{k}}\right].
\end{eqnarray}

Now assuming once again $\omega\sim k$, provided that $1\ll\frac{\w'}{k}\ll \frac{\gamma^2}{k^2}\ll A^{-1}$,  equation (\ref{a131}) turns
into equation (\ref{a109}). Consequently, we precisely obtain the
same behavior as for a perfect reflecting mirror. However, in the
case $1\ll \frac{\gamma^2}{k^2}\ll\frac{\w'}{k}\ll A^{-1}$ we observe that
\begin{eqnarray}\label{a132}
\beta^{R,R}_{\w,\w'}&\cong& \frac{\alpha}{2\pi k
i\sqrt{\w\w'}}e^{-i{\w'}/{k}}\left(i\frac{k}{\w'}\right)^{i{\w}/{k}+{1}/{2}}
\Gamma\left({1}/{2}+i{\w}/{k}\right),
\end{eqnarray}
and using the identity
$|\Gamma\left({1}/{2}+i{\w}/{k}\right)|^2=\frac{\pi}{\cosh\left(
 \pi{\w}/{k}\right)}$ (see \cite{as72}), we conclude that
\begin{eqnarray}\label{a133}
\left|\beta^{R,R}_{\w,\w'}\right|^2&\cong& \frac{1}{2\pi
k\w}\left(\frac{\gamma}{\w'}\right)^2
\left(e^{2\pi{\w}/{k}}+1\right)^{-1}.
\end{eqnarray}

Finally, a simple but rather cumbersome calculation yields in the first case
\begin{eqnarray}\label{a134}
\left|\beta_{\omega,\omega'}^{R,L}\right|^2\cong 0,
\end{eqnarray}
and in the second one
\begin{eqnarray}\label{a135}
\left|\beta_{\omega,\omega'}^{R,L}\right|^2\sim \frac{1}{\w\w'}
{\mathcal O}\left[ \left(\frac{\gamma}{\w'}\right)^2\right].
\end{eqnarray}

Then we can conclude that the number of produced particles in the mode $\w$
is approximately given by \cite{n09,eh10}

\begin{eqnarray}
 N(\w)&&\cong \int_{k}^{\gamma^2/k}\frac{1}{2\pi
k\w'}
\left(e^{2\pi{\w}/{k}}-1\right)^{-1}+\int_{\gamma^2/k}^{\infty}\frac{1}{2\pi
k\w}\left(\frac{\gamma}{\w'}\right)^2
\left(e^{2\pi{\w}/{k}}+1\right)^{-1}\nonumber\\&&=
\frac{1}{\pi k}\ln(\gamma/k)\left(e^{2\pi{\w}/{k}}-1\right)^{-1}+\frac{1}{2\pi\w}
\left(e^{2\pi{\w}/{k}}+1\right)^{-1}\cong \frac{1}{\pi k}\ln(\gamma/k)\left(e^{2\pi{\w}/{k}}-1\right)^{-1},
\end{eqnarray}
because we're assumig $k\sim \w\ll \gamma$.

That is,  the number density of produced particles in the mode $\omega$ by a partially transmitting moving mirror is finite when $u_0\rightarrow \infty$, moreover when
$\omega\sim k\ll \gamma$ the mirror radiates a thermal flux described by Bose-Einstein statistics. However when
$\omega\sim k\sim \gamma$, maybe the contribution of the sector $[\gamma^2/k,\infty)$ could be dominant and another kind of
statistics (Fermi-Dirac) would be possible. This is a situation that deserves futher investigation.

 \vspace{1cm}

\section{Vacuum fluctuations}
The re-normalization of the two point function via adiabatic regularization is given with all
the details, and the re-normalization of the stress tensor is also reviewed.
\subsection{Re-normalized two point function}

\subsubsection{Massless conformally coupled field} We start  this
lecture studying   the simplest case: a massless conformally
coupled scalar field in a Friedman-Robertson-Walker space-time.
We'll calculate the re-normalized part of the two-point function
$\langle\phi^2(\eta,\vec{x})\rangle$ using the adiabatic
regularization method. This is the simplest example, and it help
us to understand all the details of the method.

Consider the quantum scalar field
\begin{eqnarray}\label{a136}
\hat{\phi}(\eta,{\bf x})=\int_{\R^3}d^3{\bf k}\left[\hat{a}_{\bf
k}\phi_{\bf k}(\eta,{\bf x})+\hat{a}^{\dagger}_{\bf k}\phi^*_{\bf
k}(\eta,{\bf x})\right],
\end{eqnarray}
then, using the same notation as in first lecture, one has
$\chi_{\bf k}(\eta)=\frac{e^{-i |{\bf k}|\eta}}{\sqrt{2 |{\bf
k}|}}$, and thus, the two-point function is given by
\begin{eqnarray}\label{a137}
\langle\hat{\phi}^2(\eta,{\bf x})\rangle\equiv\int_{\R^3}d^3{\bf
k}|\phi_{\bf k}(\eta,{\bf
x})|^2=\frac{1}{4\pi^2C(\eta)}\int_0^{\infty}|{\bf k}|d |{\bf k}|.
\end{eqnarray}

In order to obtain the re-normalized value of the two-point
function, we'll follow  Bunch's method described in \cite{b78}:
First, one considers the adiabatic modes obtained in the WKB
approximation,
\begin{eqnarray}\label{a138}
\chi_{adi,\bf k}(\eta)=\frac{1}{\sqrt{2W_{\bf k}}}e^{-i\int W_{ \bf
k}d\eta},
\end{eqnarray}
 up to order $2$. To calculate these adiabatic modes
one has to use equation (\ref{a9}), then
 in the conformally coupled case $W_{ k}$ is given
by (see for details \cite{wi05,ap87,hmpm99})
\begin{eqnarray}\label{a139}
W_{\bf k}=\w_{\bf k}-\frac{1}{4}\frac{\w''_{\bf k}}{\w_{ \bf
k}^2}+\frac{3}{8}\frac{(\w'_{\bf k})^2}{\w_{\bf k}^3}
\end{eqnarray}
with $\w_{\bf k}^2=|{\bf k}|^2+C(\eta)m^2$, and a simple calculation
yields,
\begin{eqnarray}\label{a140}
W_{\bf k}=\w_{\bf k}-\frac{1}{8}\frac{m^2C''}{\w_{ \bf
k}^3}+\frac{5}{32}\frac{m^4(C')^2}{\w_{\bf k}^5}.
\end{eqnarray}

Once the adiabatic modes has been calculated, to obtain the
re-normalized expression of $\langle\phi^2(\eta,{\bf x})\rangle$ one
has to subtract from (\ref{a137}), the adiabatic terms up to order
two (only terms that contain, at most, two derivatives of the scalar
factor) that appear in the expression \cite{p07,mp00}
\begin{eqnarray}\label{a141}
\frac{1}{4\pi^2C(\eta)}\int_0^{\infty}\frac{|\bf k|^2}{W_{ \bf
k}}d|{\bf k}|,
\end{eqnarray}
and finally to take the limit $m\rightarrow 0$, that is:
\begin{eqnarray}\label{a142}
\langle\hat{\phi}^2(\eta,{\bf
x})\rangle_{ren}=\lim_{m\rightarrow
0}\frac{1}{4\pi^2C(\eta)}\left[\int_0^{\infty}\left(|{\bf
k}|-\frac{|{\bf k}|^2}{\w_{ \bf k}}\right)d|{\bf
k}|+\frac{m^2C''}{8}\int_0^{\infty}\frac{|{\bf k}|^2}{\w^5_{\bf
k}}d|{\bf k}|
-\frac{5m^4(C')^2}{32}\int_0^{\infty}\frac{|{\bf k}|^2}{\w^7_{ \bf
k}}d|{\bf k}| \right].\end{eqnarray}

It's not difficult to show that the final result is
\begin{eqnarray}\label{a143}
\langle\hat{\phi}^2(\eta,{\bf x})\rangle_{ren}=
-\frac{1}{96\pi^2C}\left[\frac{1}{2}\left(\frac{C'}{C}\right)^2
-\frac{C''}{C}\right]=\frac{1}{48\pi^2}\frac{a''}{a^3}=\frac{1}{288\pi^2}R,
\end{eqnarray}
which  coincides, for the de Sitter phase, with 
formula (3.19) obtained in \cite{af87}. Note also that, in the case of an universe filled by
radiation, and consequently $R=0$, $\langle\phi^2(\eta,{\bf
x})\rangle_{ren}=0$ in the massless case independently of the
coupling constant value.

\subsubsection{Massless minimally coupled field}

In this Section we consider another simple example, a massless
minimally coupled field in the flat chart of the de Sitter
space-time, where  the modes can also be  calculated exactly, they
are
\begin{eqnarray}\label{a144}
{\chi_{\bf k}(\eta)}=\left(a_{\bf k}\psi_{\bf k}(\eta)+b_{ \bf
k}\psi^*_{ \bf k}(\eta)\right),
\end{eqnarray}
where $\psi_{\bf k}(\eta)$
is given by formula (\ref{a43}) and $a_{\bf k}$ and $b_{\bf k}$
are some constants.

In general, $\langle\hat{\phi}^2(\eta,{\bf
x})\rangle=\frac{1}{2\pi^2C(\eta)}\int_0^{\infty}|{\bf
k}|^2|\chi_{\bf k}(\eta)|^2d|{\bf k}|$, has ultra-violet and
infra-red divergencies. To avoid these last ones (see for details
\cite{fp77,vf82}), we consider a transition from the radiation
dominated phase to the de Sitter one, described by the following
scale factor:
\begin{eqnarray}\label{a145}
a(\eta)=\left\{\begin{array}{cc} 2-\eta/\eta_0&\eta<\eta_0\\
\eta_0/\eta&\eta>\eta_0,\end{array}\right.
\end{eqnarray}
with $\eta_0=-1/H$.

The modes $\chi_{\bf k}$ for $\eta<\eta_0$, are given by
${e^{-i|{\bf k}|\eta}}$. Note that these modes correspond to the
usual choice of the vacuum state for a massless field in the
radiation phase, because in that phase the scalar curvature
vanishes, and then $\chi_{\bf k}$ satisfy the equation $\chi_{\bf
k}''+|{\bf k}|^2\chi_{\bf k}=0$, and consequently, the vacuum state
is obtained in the same way as in the Minkowskian case, that is,
from the modes $e^{-i|{\bf k}|\eta}$.

 Matching at the point
$\eta=\eta_0$ the modes an their derivatives, one obtains
\begin{eqnarray}\label{a146}
a_{\bf k}=1+\frac{H}{i|{\bf k}|}-\frac{H^2}{2|{\bf k}|^2},\qquad
b_{\bf k}=-\frac{H^2}{2|{\bf k}|^2}e^{\frac{2i|{\bf k}|}{H}}=a_{\bf
k}+\frac{2i|{\bf k}|}{3H}+{\mathcal O}\left(|{\bf k}|^2/H^2\right).
\end{eqnarray}

With these coefficients, for small values of $|{\bf k}|$, in the de
Sitter phase $(\eta>\eta_0)$, one has
\begin{eqnarray}\label{a147}
|\chi_{\bf k}|^2=\frac{1}{{2|{\bf
k}|}}\left[\left(\frac{2}{3H\eta}+2+\frac{H^2\eta^2}{6}\right)^2+{\mathcal
O}\left(|{\bf k}|^2/H^2\right)\right],
\end{eqnarray}
what shows that there is not infra-red divergencies.

To analyze ultra-violet divergencies we calculate for large $|{\bf
k}|$
\begin{eqnarray}\label{a148}
|\chi_{\bf k}|^2=\frac{1}{{2|{\bf k}|}}\left[1+\frac{1}{|{\bf
k}|^2\eta^2}-\frac{H^2}{|{\bf k}|^2}\cos(2|{\bf
k}|(H^{-1}+\eta)+{\mathcal O}\left(H^3/|{\bf k}|^3\right)\right],
\end{eqnarray}
what shows that the terms that give ultra-violet divergencies in
$\langle\hat{\phi}^2(\eta,{\bf x})\rangle$ are
\begin{eqnarray}\label{a149}
\frac{\eta^2H^2}{4\pi^2}\int^{\infty}|{\bf k}|d|{\bf k}|\quad
\mbox{and} \quad \frac{H^2}{4\pi^2}\int^{\infty}\frac{1}{|{\bf
k}|}d|{\bf k}|.
\end{eqnarray}

Once we have separated the divergent terms, we calculate, up to
order $2$, the adiabatic terms that  for $\eta
>\eta_0$, are given by
\begin{eqnarray}\label{a150}
W_{\bf k}=\w_{\bf k}-\frac{1}{\eta^2\w_{\bf
k}}-\frac{1}{8}\frac{m^2C''}{\w_{\bf
k}^3}+\frac{5}{32}\frac{m^4(C')^2}{\w_{\bf k}^5}.
\end{eqnarray}

From  that, one sees that the divergent parts of (\ref{a141}) are,
in the de Sitter phase, given by
\begin{eqnarray}\label{a151}
\frac{\eta^2H^2}{4\pi^2}\int^{\infty}\frac{|{\bf k}|^2}{\w_{\bf
k}}d|{\bf k}|\quad \mbox{and} \quad
\frac{H^2}{4\pi^2}\int^{\infty}\frac{|{\bf k}|^2}{\w_{\bf
k}^3}d|{\bf k}|.
\end{eqnarray}

Subtracting, for large frequencies,  the divergent part of
(\ref{a141}) from (\ref{a149}), for example for $|{\bf
k}|>|\eta|^{-1}$, one gets
\begin{eqnarray}\label{a152}
\lim_{m\rightarrow
0}\frac{\eta^2H^2}{4\pi^2}\int_{|\eta|^{-1}}^{\infty}\left(|{\bf
k}|-\frac{|{\bf k}|^2}{\w_k}\right)d|{\bf k}|=0\quad \mbox{and}
\quad \lim_{m\rightarrow
0}\frac{H^2}{4\pi^2}\int_{|\eta|^{-1}}^{\infty}\left(\frac{1}{|{\bf
k}|}-\frac{|{\bf k}|^2}{\w_{\bf k}^3}d|{\bf k}|\right)=0,
\end{eqnarray}
what  shows that the ultra-violet divergencies are canceled.

The problem now is that the subtracted adiabatic term
$\frac{H^2}{4\pi^2}\int_0^{H}\frac{|{\bf k}|^2}{\w_{\bf k}^3}d|{\bf
k}|$ contains an infra-red divergency, because it diverges when the
mass approaches to zero. It is important to remark that this term
can be written as
$-\frac{R}{8\pi^2}\int_0^{H}\left(\xi-\frac{1}{6}\right)\frac{|{\bf
k}|^2}{\w_{\bf k}^3}d|{\bf k}|$, that is, it does not appear in the
conformal coupled case.

The solution of this infra-red divergency emerges from the
following observation: The adiabatic approximation is based on
modes of the form (\ref{a138}), and it is clear that this form
only has sense if the exact modes, namely $\chi_{\bf k}$, are
oscillating, that is, if $\chi_{\bf k}$ satisfy the equation
(\ref{a2}), with
\begin{eqnarray}\label{a153}
\Omega^2_{\bf k}(\eta)>0.
\end{eqnarray}

In our case this conditions means $|{\bf
k}|>\sqrt{2}|\eta^{-1}|=\sqrt{2}He^{Ht}$, and thus, we  must only
subtract  adiabatic modes well inside in the Hubble horizon at
time $t$. Consequently, no infra-red divergencies appears.

Here an important remark is in order: Our recipe to eliminate the infra-red  divergency does not
affect  to the conservation of the renormalized stress tensor
which for a FRW metric reduces to  $(\rho C^{3/2})'+p(C^{3/2})'=0$, where $\rho$ is the energy density and
$p$ is the pressure, because the adiabatic regularization
lies in subtracting adiabatic terms up to a given order and then the conservation equation is
safisfied for each order, and more important, this subtraction
can be performed mode by mode \cite{pf74, b80}. This means that if  one denotes by
$\rho_{ad}(\phi_k)$ and
$p_{ad}(\phi_k)$ the adiabatic terms of the energy density and pressure for
the adiabatic mode defined in equation (\ref{a138}), 
 the conservation equation $(\rho_{ad}(\phi_k) C^{3/2})'+p_{ad}(\phi_k)(C^{3/2})'=0$ will be satisfied, and since the substraction is performed mode by mode,
 one can subtracts a given number of modes mantaining the conservation equation.

Summarizing, the re-normalized quantity is given by
\begin{eqnarray}\label{a154}
\langle\hat{\phi}^2(\eta,{\bf x})\rangle_{ren}=\lim_{m\rightarrow
0}\frac{1}{2\pi^2C(\eta)}\left(\int_0^{\infty}|{\bf
k}|^2|\chi_{\bf k}(\eta)|^2d|{\bf k}|-
\frac{1}{2}\int_{\sqrt{2}/|\eta|}^{\infty}\frac{|{\bf
k}|^2}{W_{\bf k}}d|{\bf k}|\right).
\end{eqnarray}

Now since, the ultraviolet divergencies are cancelled, and
\begin{eqnarray}\label{a155}
\lim_{m\rightarrow 0}\int_{\sqrt{2}/|\eta|}^{\infty}\frac{m^2|{\bf
k}|^2}{\w_{\bf k}^5}d|{\bf k}|=\lim_{m\rightarrow
0}\int_{\sqrt{2}/|\eta|}^{\infty}\frac{m^4|{\bf k}|^2}{\w_{\bf
k}^7}d|{\bf k}|=0,
\end{eqnarray}
using (\ref{a148}) we finally get
\begin{eqnarray}\label{a156}
\langle\hat{\phi}^2(\eta,{\bf
x})\rangle_{ren}&&=\frac{\eta^2H^2}{2\pi^2}\int_0^{\sqrt{2}/|\eta|}|{\bf
k}|^2|\chi_{\bf k}(\eta)|^2d|{\bf k}|
\nonumber\\&&+
\frac{\eta^2H^2}{4\pi^2}\int_{\sqrt{2}/|\eta|}^{\infty}\left(-\frac{H^2}{|{\bf
k}|^2}\cos(2|{\bf k}|(H^{-1}+\eta)+{\mathcal O}\left(H^3/|{\bf
k}|^3\right)\right)|{\bf k}|d|{\bf k}|.
\end{eqnarray}

If we are interested in the late time behavior ($|\eta|H\ll 1$, i.e.,
$Ht\gg 1$),  we can make the following approximation
\begin{eqnarray}\label{a157}
\langle\hat{\phi}^2(\eta,{\bf x})\rangle_{ren}\cong\frac{\eta^2
H^2}{2\pi^2}\int_0^{\sqrt{2}/|\eta|}|{\bf k}|^2|\chi_{\bf
k}(\eta)|^2d|{\bf k}|\cong \frac{1}{9\pi^2}\int_0^H{|{\bf
k}|}d|{\bf
k}|+\frac{H^2}{4\pi^2}\int_H^{\sqrt{2}/|\eta|}\frac{1}{|{\bf
k}|}d|{\bf k}|,
\end{eqnarray}
where, in the first integral we have used the approximation
$|\chi_{\bf k}(\eta)|^2\cong\frac{2}{9\eta^2H^2|{\bf k}|}$ (eq.
(\ref{a147})), and in the second one $|\chi_{\bf
k}(\eta)|^2\cong\frac{1}{2\eta^2|{\bf k}|^3}$ (eq. (\ref{a148})).

Finally, after integration,  at late times we get
\begin{eqnarray}\label{a158}
\langle\hat{\phi}^2(\eta,{\bf
x})\rangle_{ren}\cong\frac{H^2}{18\pi^2}+\frac{H^2}{4\pi^2}(\frac{1}{2}\ln
2+Ht)\cong\frac{H^3}{4\pi^2}t,
\end{eqnarray}
that coincides with the early result obtained in
\cite{l82,s82,vf82}.

Note also that formula (\ref{a153}) justify the prescription given
in \cite{v83, h92}, where the authors assumes that only modes
outside to the Hubble horizon  at time $t$ contribute to the value
of the re-normalized two-point function.

In the opposite case, that is, for a few Hubble times, (for
example $t=1/H$), it is not difficult to show that
$\langle\hat{\phi}^2(\eta,{\bf x})\rangle_{ren}\sim {\mathcal
O}(H^2)$.

\begin{remark}
In inflationary cosmology the re-normalization of the two point
function is sometimes obtained in a different way (see for example
\cite{li05}). The modes $\chi_{\bf k}(\eta)$ are given by formula
$(\ref{a43})$, and thus
$$
\langle\hat{\phi}^2(\eta,{\bf
x})\rangle=\frac{1}{4\pi^2C(\eta)}\int_0^{\infty} |{\bf
k}|\left(1+\frac{1}{|{\bf k}|^2\eta^2}\right)d|{\bf k}|.
$$

To avoid the infra-red divergency one assumes that the initial
size of the universe is of the order $H^{-1}$, then the Fourier
expansion  shows the modes has to be  wave-length  smaller than
the Hubble horizon, that is, impose the cut-off $|{\bf k}|\geq H$.
And to avoid the ultra-violet divergency one has to subtract
adiabatic modes that satisfy equation $(\ref{a153})$, i.e.,
adiabatic modes well inside in the Hubble horizon, and
consequently one only has to take into account the modes that
leave the horizon, more precisely modes that satisfy $H\leq |{\bf
k}|\leq \sqrt{2}/|\eta|$,
 then from
$(\ref{a152})$ and $(\ref{a155})$ one has
$$
\langle\hat{\phi}^2(\eta,{\bf
x})\rangle=\frac{1}{4\pi^2C(\eta)}\int_H^{\sqrt{2}/|\eta|} |{\bf
k}|\left(1+\frac{1}{|{\bf k}|^2\eta^2}\right)d|{\bf k}|.
$$

Finally note that the first term is the usual contribution from
vacuum fluctuations in Minkowski space and must be eliminated by
re-normalization, which give the following re-normalized two point
function
$$
\langle\hat{\phi}^2(\eta,{\bf
x})\rangle_{ren}=\frac{1}{4\pi^2C(\eta)}\int_H^{\sqrt{2}/|\eta|}
\frac{1}{|{\bf k}|\eta^2}d|{\bf
k}|=\frac{H^2}{4\pi^2}(\frac{1}{2}\ln
2+Ht)\cong\frac{H^3t}{4\pi^2},
$$
when $Ht\geq 1.$
\end{remark}

\subsubsection{Massive case}

First, we study the minimally coupled case with $m\ll H$ in the de
Sitter phase. (This situation appears when the inflation fields is
in the slow-roll phase, and  scalar field fluctuations described
by the two-point function are very important in order to
understand the self-reproducing universes in inflationary
cosmology (see for example \cite{l82, v83})). In  last Section, we
have seen that the re-normalized two-point function is given by
the formula (\ref{a154}) without the limit $m\rightarrow 0$,
because the adiabatic modes that we have to subtract satisfy
$|{\bf
k}|>\frac{\sqrt{2}}{|\eta|}\sqrt{1-\frac{m^2}{2H^2}}\cong\frac{\sqrt{2}}{|\eta|}$.

The calculation of the  two-point in the massive case is more
difficult than in the massless one, however,  at late times, it is
possible  to approximate its behavior very well. To do this, note
that the adiabatic regularization method guarantees that
\begin{eqnarray}\label{a159}
\frac{1}{2\pi^2 C(\eta)}\int_{\sqrt{2}/|\eta|}^{\infty}|{\bf
k}|^2\left(|\chi_{\bf k}(\eta)|^2- \frac{1}{2W_{\bf
k}}\right)d|{\bf k}|,
\end{eqnarray}
is convergent \cite{pf74}. To perform this integral one can choose
mode solutions that correspond to the Bunch-Davies vacuum state,
i.e., $\chi_{\bf
k}(\eta)=\sqrt{\frac{\pi\eta}{4}}e^{-i(\frac{\pi\nu}{2}+\frac{\pi}{4})}H_{\nu}^{(2)}(|{\bf
k}|\eta)$ with $\nu\equiv \sqrt{\frac{9}{4}-\frac{m^2}{H^2}}\cong
\frac{3}{2}-\frac{m^2}{3H^2}$. Then, at late times, using the
asymptotic expansion for large arguments of the Hankel functions
(formulae 9.2.8-9.2.10 of \cite{as72}) one can shows that the
divergent terms of (\ref{a159}) exactly cancels, and since a easy
calculation proves that the convergent ones are of the order
${\mathcal O}(m^2)$, one can disregard its contribution to the
two-point function. Thus, at late times, we have
\begin{eqnarray}\label{a160}\langle\hat{\phi}^2(\eta,{\bf x})\rangle_{ren}\cong
\frac{\eta^2 H^2}{2\pi^2}\int_{0}^{H}|{\bf k}|^2|\chi_{\bf
k}(\eta)|^2d|{\bf k}|+\frac{\eta^2
H^2}{2\pi^2}\int_H^{\sqrt{2}/|\eta|}|{\bf k}|^2|\chi_{\bf
k}(\eta)|^2d|{\bf k}|.
\end{eqnarray}

In order to avoid infra-red divergency, we can calculate the first
integral assuming a phase transition to the radiation dominated
universe to a de Sitter phase at time $\eta_0=-1/H$.
In fact, we consider the general mode
solutions in the de Sitter phase
\begin{eqnarray}\label{a161}
\chi_{\bf k}(\eta)=\sqrt{\pi/4}\eta^{1/2}\left(a_{\bf
k}H_{\nu}^{(2)}(|{\bf k}|\eta)+b_{\bf k}H_{\nu}^{(1)}(|{\bf
k}|\eta)\right),
\end{eqnarray}
and match  the modes and their temporal derivatives at
$\eta_0=-1/H$, to obtain
\begin{eqnarray}\label{a162}
a_{\bf k}=\frac{1}{2i}\sqrt{\frac{\pi
|{\bf k}|\eta_0}{2}}\left(\left(-i+\frac{H}{2|{\bf k}|}\right)
H_{\nu}^{(1)}(|{\bf k}|\eta_0)-{H_{\nu}^{(1)}}'(|{\bf k}|\eta_0)\right)e^{i|{\bf k}|/H}\nonumber\\
b_{\bf k}=-\frac{1}{2i}\sqrt{\frac{\pi |{\bf
k}|\eta_0}{2}}\left(\left(-i+\frac{H}{2|{\bf
k}|}\right)H_{\nu}^{(2)}(|{\bf k}|\eta_0)-{H_{\nu}^{(2)}}'(|{\bf
k}|\eta_0)\right)e^{i|{\bf k}|/H}.
\end{eqnarray}

Then from
the small-argument limit
\begin{eqnarray}\label{a163}
H_{\nu}^{(2)}(|{\bf
k}|\eta_0)\cong-H_{\nu}^{(1)}(|{\bf k}|\eta_0)\cong
\frac{i}{\pi}\Gamma(\nu)\left(\frac{|{\bf
k}|\eta_0}{2}\right)^{-\nu},
\end{eqnarray}
and using that (see \cite{as72})
\begin{eqnarray}\label{a164}
{H_{\nu}^{(1,2)}}'(z)=H_{\nu-1}^{(1,2)}(z)-\frac{\nu}{z}H_{\nu}^{(1,2)}(z),
\end{eqnarray}
 one arrives at the result $|\chi_{\bf
k}(\eta)|^2\cong\frac{2}{9|{\bf k}|}(H|\eta|)^{1-2\nu}$, and
finally one obtains
\begin{eqnarray}\label{a165}
\frac{\eta^2H^2}{2\pi^2}\int_{0}^{H}|{\bf k}|^2|\chi_{\bf
k}(\eta)|^2d|{\bf k}|\cong
\frac{H^2}{18\pi^2}e^{-\frac{2m^2t}{3H}}\cong
\left\{\begin{array}{ccc}\frac{H^2}{18\pi^2}&\mbox{when}&1/H\ll
t\ll H/m^2
\\ &&\\0&\mbox{when}&t\gg H/m^2.\end{array}\right.
\end{eqnarray}
with agrees with the first term of the right hand side of
(\ref{a158}).

The second one, can be done using the following approximation, valid
of $|{\bf k}|>H$,
\begin{eqnarray}\label{a166}
\chi_{\bf
k}(\eta)\cong\sqrt{\frac{\pi\eta}{4}}e^{-i(\frac{\pi\nu}{2}+\frac{\pi}{4})}H_{\nu}^{(2)}(|{\bf
k}|\eta).
\end{eqnarray}
Effectively, from (\ref{a162}) one can easily obtains, in the
range $|{\bf k}|>H$, $a_{\bf k}\cong 1$ and $b_{\bf k}\cong 0$.
Then, since at late time we have $H_{\nu}^{(2)}(|{\bf
k}|\eta)\cong-\frac{i}{\pi}\Gamma(\nu)\left(\frac{|{\bf
k}|\eta}{2}\right)^{-\nu}$, inserting this expression in the
second integral of (\ref{a160}), one obtains, for a massive
minimally coupled field with $m\ll H$,
\begin{eqnarray}\label{a167}
\frac{\eta^2H^2}{2\pi^2}\int_H^{\sqrt{2}/|\eta|}|{\bf
k}|^2|\chi_{\bf k}(\eta)|^2dk\cong\frac{3H^4}{8m^2\pi^2}\left[
2^{\frac{m^2}{3H^2}}-(H|\eta|)^{\frac{2m^2}{3H^2}}\right]\cong
\frac{3H^4}{8m^2\pi^2}\left[ 1-e^{-\frac{2m^2t}{3H}}\right],
\end{eqnarray}
because we have assumed $m\ll H$.

Then, since the first integral in the right hand side of
(\ref{a160}) is smaller than the second one, depending on the
value of $m^2t/H$ one has
\begin{eqnarray}\label{a168}\langle\hat{\phi}^2(\eta,{\bf x})\rangle_{ren}\cong
\frac{3H^4}{8m^2\pi^2}\left[ 1-e^{-\frac{2m^2t}{3H}}\right]\cong
\left\{\begin{array}{ccc}\frac{H^3}{4\pi^2}t&\mbox{when}&1/H\ll
t\ll H/m^2\\
&&\\\frac{3H^4}{8m^2\pi^2}&\mbox{when}&t\gg
H/m^2,\end{array}\right.
\end{eqnarray}
which demonstrates, at late times, the formula $(7)$ of \cite{l82}.

\begin{remark}
In inflationary cosmology one can chooses $\chi_{\bf
k}(\eta)$ given in formula (\ref{a166}), then
if inflation starts at $t=0$ and the initial size of the universe
is of the order $1/H$, after subtracting the adiabatic modes well
inside in the Hubble horizon at time $t$, one only has to take
into account the modes that leave the Hubble horizon. Finally one
can uses the small-argument limit (\ref{a164}), which is
equivalent to eliminate the Minkowskian vacuum fluctuations, to
get
\begin{eqnarray*}
 \langle\hat{\phi}^2(\eta,{\bf
x})\rangle_{ren}\cong\frac{3H^4}{8m^2\pi^2}\left[
2^{\frac{m^2}{3H^2}}-(H|\eta|)^{\frac{2m^2}{3H^2}}\right]\cong
\frac{3H^4}{8m^2\pi^2}\left[ 1-e^{-\frac{2m^2t}{3H}}\right],
\end{eqnarray*}
because we are assuming $m\ll H$.
\end{remark}

\vspace{0.5cm}

Finally we calculate the re-normalized two-point function,  for a
massive conformally coupled field with $m\ll H$ in the de Sitter
phase, given by
\begin{eqnarray}\label{a169}
\langle\hat{\phi}^2(\eta,{\bf
x})\rangle_{ren}=\frac{\eta^2H^2}{2\pi^2}\left(\int_0^{\infty}|{\bf
k}|^2|\chi_{\bf k}(\eta)|^2d|{\bf k}|-
\frac{1}{2}\int_{0}^{\infty}\frac{|{\bf k}|^2}{W_{\bf k}}d|{\bf
k}|\right),
\end{eqnarray}
where $W_{\bf k}$ is given by formula (\ref{a140}). (Note that, in
the conformally coupled case we do not have to disregard any
adiabatic mode.)

At late times, with the same kind of argument used to disregard the
 terms that appear in equation (\ref{a159}), we can do the following
approximation
\begin{eqnarray}\label{a170}
\langle\hat{\phi}^2(\eta,{\bf
x})\rangle_{ren}\cong\frac{\eta^2H^2}{2\pi^2}\left(\int_0^{A|\eta|^{-1}}|{\bf
k}|^2|\chi_{\bf k}(\eta)|^2d|{\bf k}|-
\frac{1}{2}\int_{0}^{A|\eta|^{-1}}\frac{|{\bf k}|^2}{W_{\bf
k}}d|{\bf k}|\right),
\end{eqnarray}
where $A$ is some dimensionless constant of order $1$. Actually, one can chooses $A=1$, and the reasoning
does not change, because $|\eta|^{-1}$ is large enough at late times.

Since in this case there is not infra-red divergency,  we can
choose the modes corresponding to the Bunch-Davies vacuum state,
that is, $\chi_{\bf
k}(\eta)=\sqrt{\frac{\pi\eta}{4}}e^{-i(\frac{\pi\nu}{2}+\frac{\pi}{4})}H_{\nu}^{(2)}(|{\bf
k}|\eta)$. Then for $k<A/|\eta|$, a good approximation is given by
$\chi_{\bf k}(\eta)\cong\frac{i}{\sqrt{2|{\bf k}|}}$, and since
\begin{eqnarray}\label{a171}
\frac{H^2\eta^2}{4\pi^2}\int_0^{A|\eta|^{-1}}\left(|{\bf
k}|-\frac{|{\bf k}|^2}{\w_{\bf k}}\right)d|{\bf k}|\sim {\mathcal
O}\left(m^2\ln\left(H/m\right)\right),
\end{eqnarray}
one can disregard this term (it is small compared with the other
ones that are of order ${\mathcal O}(H^2)$) and, at late times, we
get the same result obtained in formula (\ref{a143}) with
$R=12H^2$, that is,
\begin{eqnarray}\label{a172}
\langle\hat{\phi}^2(\eta,{\bf x})\rangle_{ren}=
\frac{H^2}{24\pi^2},
\end{eqnarray}
because the  leading terms  in (\ref{a170}) are the same as in
(\ref{a142}).

\subsection{Re-normalized stress-tensor}
In this section we review some classic results about the re-normalization of the stress-tensor
in FRW cosmologies.

The vacuum energy density $\rho$ is given by

\begin{eqnarray}\label{b173}
\rho\equiv\langle\hat{T}_{tt}\rangle=(4\pi^2C^2)^{-1}\int_0^{\infty}d|{\bf
k}||{\bf k}|^2\left\{(|\chi'_{\bf k}|^2+\omega^2_{\bf k}|\chi_{\bf
k}|^2)+3(\xi-1/6)\left[D(\chi_{\bf
k}\chi^{*'}_{\bf k}+\chi^*_{\bf k}\chi'_{\bf
k})-\frac{1}{2}D^2|\chi_{\bf k}|^2\right]\right\},
\end{eqnarray}
where $D=C'/C$.

To obtain the re-normalized value, one can uses the adiabatic regularization which consist in subtracting adiabatic modes
up to order four. Following \cite{b80} one has to subtract the
following divergent terms:
\begin{eqnarray}\label{b174}
(4\pi^2C^2)^{-1}\int_0^{\infty}d|{\bf k}||{\bf k}|^2\omega_{\bf
k},\quad \frac{3(\xi-1/6)D^2}{16\pi^2C^2}\int_0^{\infty}d|{\bf
k}|\frac{|{\bf k}|^2}{\omega_{\bf
k}}-\frac{3(\xi-1/6)D^2m^2}{16\pi^2C}\int_0^{\infty}d|{\bf
k}|\frac{|{\bf k}|^2}{\omega^3_{\bf k}}
\end{eqnarray}
and
\begin{eqnarray}\label{b175}
-\frac{(\xi-1/6)^2}{128\pi^2C^2} (72D''D-36D'^2-27D^4)
\int_0^{\infty}d|{\bf k}|\frac{|{\bf k}|^2}{\omega^3_{\bf k}},
\end{eqnarray}
and the finite terms:
\begin{eqnarray}\label{b176}
\frac{m^2D^2}{384\pi^2C}
-&&\frac{1}{2880\pi^2C^2}\left(\frac{3}{2}D''D-\frac{3}{4}D'^2-\frac{3}{8}D^4\right)\nonumber
 \\&&+\frac{(\xi-1/6)}{256\pi^2C^2}
(8D''D-4D'^2-3D^4)+\frac{(\xi-1/6)^2}{64\pi^2C^2} (18D'D^2+9D^4).
\end{eqnarray}

The simplest example is when one considers a massless minimally
coupled scalar field, in that case the modes are $\chi_{\bf k}(\eta)=\frac{e^{-i |{\bf k}|\eta}}{\sqrt{2 |{\bf
k}|}}$
then one easily obtains \cite{b78,dfcb77,bd77}
\begin{eqnarray}\label{b177}
\rho_{vac}\equiv\langle\hat{T}_{tt}\rangle_{ren}=\frac{1}{2880\pi^2C^2}\left(\frac{3}{2}D''D-\frac{3}{4}D'^2-\frac{3}{8}D^4\right).
\end{eqnarray}

In the flat chart of the de-Sitter space-time choosing the modes
\begin{eqnarray}\label{b178}
\chi_{\bf k}(\eta)=C\sqrt{\frac{\pi\eta}{4}}H^{(2)}_{\nu}(|{\bf
k}| \eta),
\end{eqnarray}
with $\nu\equiv\sqrt{\frac{9}{4}-\frac{m^2}{H^2}-12\xi}$ and
$C\equiv e^{-i(\frac{\pi\nu}{2}+\frac{\pi}{4})}$, Bunch and Davies
obtained in \cite{bd78} using point-splitting regularization (which for scalar massive field is equivalent to adiabatic regularization
\cite{b80,bi78, bcf78})
\begin{eqnarray}\label{b179}
\rho_{vac}\equiv\langle\hat{T}_{tt}\rangle_{ren}=&&\frac{1}{64\pi^2}\left\{m^2\left[
m^2+(12\xi-2)H^2\right]\left[\psi\left(\frac{3}{2}+\nu\right)+\psi\left(\frac{3}{2}-\nu\right)-\ln\left(\frac{m^2}{H^2}\right)\right]
\nonumber\right.\\&&\left.-m^2(12\xi-2)H^2-\frac{2}{3}m^2H^2-\frac{1}{2}(12\xi-2)^2H^4+\frac{H^4}{15}\right\},
\end{eqnarray}
where $\psi$ denotes the digamma function. (In exact agreement with the previous resolt obtained in \cite{dc76}
using Schwinger-DeWitt regularization procedure \cite{dw75})

This result holds for all values of $m$ and $\xi$ except in the
massless minimally coupled case, because when $m$ and $\xi$ are
close to zero one has \cite{kg93}
\begin{eqnarray}\label{b180}
\frac{1}{64\pi^2}m^2\left[
m^2+(12\xi-2)H^2\right]\psi\left(\frac{3}{2}-\nu\right)\cong\frac{1}{128\pi^2}\frac{H^2}{1+\frac{12\xi
H^2}{m^2 }},
\end{eqnarray}
and thus the limit in (\ref{b179}) gives different answers
depending the way that $m$ and $\xi$ approach to the origin.
In order to calculate $\rho_{vac}$ in the massless minimally coupled case using adiabatic regularization on has to
consider the $\xi=0$, use the modes given in (\ref{a43}) and finally take $m\rightarrow 0$ \cite{b78,b80}, gettin 
\begin{eqnarray}\label{b181}
\rho_{vac}=-\frac{29H^4}{960\pi^2},
\end{eqnarray}
which can be also obtained taking $\xi=0$ in (\ref{b179}) and afterwards $m\rightarrow 0$.

\begin{remark}
 It's well-known that the trace of the stress tensor corresponding
to a massless conformally coupled field is zero, however after
regularization its value is not necessarily zero. It was showed in
\cite{b80} that the trace anomaly provided by adiabatic
regularization, in the massless conformally coupled case is given
by
\begin{eqnarray}\label{b182}
T_{vac}\equiv\langle T^{\alpha}_{\alpha}\rangle_{ren}=
\frac{1}{960\pi^2C^2}(D'''-D'D^2).
\end{eqnarray}
Then from the relation $T_{vac}=\rho_{vac}-3p_{vac}$ one gets the
value of the vacuum pressure $p_{vac}\equiv\langle
T_{xx}\rangle_{ren}$. In fact, if one knows one of these values
the others come from the trace anomaly relation
$T_{vac}=\rho_{vac}-3p_{vac}$ and the conservation equation
$\left(\rho_{vac}C^{3/2}\right)'+p_{vac}\left(C^{3/2}\right)'=0$.
\end{remark}

\vspace{1cm}

\section{Avoidance of cosmological singularities}
The vacuum quantum effects due to a massless conformally couped field are taken
into account in order to avoid classical cosmological singularities.

\subsection{Review of classical and quantum cosmology}

{\bf Classical cosmology:}
We will use the following
notation: $\kappa^2=16\pi G=16\pi/m_p^2$, being $G$ Newton's constant, $\rho$
energy density, $p$  pressure, $\omega$  a dimensionless parameter
and $H=\dot{a}/a$  the Hubble parameter, being $a$ the scale factor and the dot denotes
denotes the derivative with respect to the cosmic time $t$.
Then the Friedmann equation and conservation equation, for a flat
Friedmann-Robertson-Walker cosmology, can be written respectively
as:
\begin{eqnarray}\label{a183}
H^2=\frac{\kappa^2}{6}\rho,\quad\dot{\rho}=-3H(\rho+p),
\end{eqnarray}
and the equation of state for a barotropic perfect fluid, that we
will consider in the paper,  has the form $p=\omega\rho.$

With the derivative of the Friedmann equation, and the other two
equation one easily obtains the  ``acceleration'' equation
\begin{eqnarray}\label{a184}
\dot{H}=-\frac{\kappa^2}{4}(1+\omega)\rho.
\end{eqnarray}

Combining (\ref{a183}) and (\ref{a184}), one can deletes $\rho$, and
obtains the equation
$\dot{H}=-\frac{3}{2}(1+\omega)H^2,$ then
integrating one gets
\begin{eqnarray}\label{a185}
{H}(t)=\frac{2}{3(1+\omega)}\frac{1}{t-t_s},
\end{eqnarray}
where $t_s\equiv t_0-\frac{2}{3H_0(1+\omega)}$, being $H_0=H(t_0)$
the initial condition.

From the definition of the Hubble parameter,  the following behavior
for the scale factor is obtained
\begin{eqnarray}\label{a186}
a(t)=a_0\left(\frac{t-t_s}{t_0-t_s}\right)^{\frac{2}{3(1+\omega)}},
\end{eqnarray}
and from the Friedmann equation, one has
\begin{eqnarray}\label{a187}
\rho(t)=\frac{8}{3\kappa^2(1+\omega)^2}\frac{1}{(t-t_s)^2}.
\end{eqnarray}

The following remark is in order: if one assumes $H_0>0$, then for
$\omega>-1$ one has $t_s<t_0$, that is, the singularity is at early
times (Big Bang singularity), on the other hands, for $\omega<-1$
one has $t_s>t_0$, that is, the singularity is at late times (Big
Rip singularity) \cite{ckw03}.

\vspace{1cm}

{\bf Quantum Effects:} Using the same notation as \cite{d77}, for a
massless field conformally coupled with gravity, one has the
following expression for the
 trace anomaly
\begin{eqnarray}\label{a188}
T_{vac}=6\alpha(\dddot{H}+12H^2\dot{H}+7H\ddot{H}+4\dot{H}^2)-12\beta
(H^4+H^2\dot{H})
\end{eqnarray}
with (see for example \cite{no04})
\begin{eqnarray}\label{a189}
\alpha=\frac{1}{2880\pi^2}(N_0+6N_{1/2}+12N_1),\quad
\beta=\frac{-1}{2880\pi^2}(N_0+11N_{1/2}+62N_1)
\end{eqnarray}
where $N_0$ is the number of scalar fields, $N_{1/2}$ is the number
of four components neutrinos and $N_1$ is the number of
electromagnetic fields.

\begin{remark}
The constants $\alpha$ and $\beta$ come from the regularization
process. For example dimensional regularization gives formula
(\ref{a189}), and point-splitting gives (see \cite{d77})
$\alpha=\frac{1}{2880\pi^2}(N_0+3N_{1/2}-18N_1)$,
$\beta=\frac{-1}{2880\pi^2}(N_0+11/2N_{1/2}+62N_1)$. Then, since the
method of regularization influences these values, and it is
uncertain what fields are present in our universe, one can considers
all values of both parameters.
\end{remark}

\begin{remark}
 The relation between the notation of this section and the one of \cite{v85} that we
used in section (II.A.3) is
$$M^2=-\frac{2}{\kappa^2 \alpha},\qquad H_0^2=-\frac{2}{\kappa^2 \beta}.$$
\end{remark}

As explained in Remark III.1,
to obtain the vacuum energy density one can uses the trace anomaly
$T_{vac}=\rho_{vac}-3p_{vac}$, and the conservation equation. The
result is
\begin{eqnarray}\label{a190}
\rho_{vac}=6\alpha(3H^2\dot{H}+H\ddot{H}-\frac{1}{2}\dot{H}^2)-3\beta
H^4,
\end{eqnarray}
which coincides with eq. (\ref{b177}) if one only considers scalar fields.
Then taking into account this vacuum energy density, the modified
Friedmann equation behaves
\begin{eqnarray}\label{a191}
H^2=\frac{\kappa^2}{6}(\rho+\rho_{vac}).
\end{eqnarray}

With the derivative of this last equation, the conservation equation
and the trace anomaly, one obtains the modified acceleration
equation
\begin{eqnarray}\label{a192}
\dot{H}=-\frac{\kappa^2}{4}\left((1+\omega)\rho+\rho_{vac}+\frac{1}{3}(\rho_{vac}-T_{vac})\right).
\end{eqnarray}

From both equations, one can deletes $\rho$, and one obtains the
following third order differential equation
\begin{eqnarray}\label{a193}
-\frac{4}{\kappa^2}\dot{H}-\rho_{vac}-\frac{1}{3}(\rho_{vac}-T_{vac})=(1+\omega)\frac{6}{\kappa^2}H^2-(1+\omega)\rho_{vac},
\end{eqnarray}
that in terms of the Hubble parameter is given by
\begin{eqnarray}\label{a194}
-\frac{4}{\kappa^2}\dot{H}-(1+\omega)\frac{6}{\kappa^2}H^2&-&3\beta(\omega+1)H^4+(18\alpha(\omega+1)-4\beta)H^2\dot{H}
\nonumber\\&&
+6\alpha(\omega+2)H\ddot{H}+3\alpha(\omega+3)\dot{H}^2+2\alpha\dddot{H}=0.
\end{eqnarray}

\subsection{$\alpha=0$}
In this section we consider the simplest case ($\alpha=0$), and we
will see that,  quantum effects don't avoid the singularities.
Equation (\ref{a12}) reduces to the following first order
differential equation
\begin{eqnarray}\label{a195}
\dot{H}=-\frac{1+\omega}{4} \left(\frac{\frac{6}{\kappa^2}H^2+3\beta
H^4}
 {\frac{1}{\kappa^2}+\beta H^2}\right).
\end{eqnarray}

\begin{enumerate}\item
First, we consider the case $\alpha=0$ and $\beta>0$. Integrating
equation (\ref{a13}) one obtains
\begin{eqnarray}\label{a196}
-1/H(t)+\sqrt{\frac{\beta\kappa^2}{2}}\arctan\left(\sqrt{\frac{\beta\kappa^2}{2}}H(t)\right)=
-\frac{3}{2}(1+\omega)(t-t_s)+
\sqrt{\frac{\beta\kappa^2}{2}}\arctan\left(\sqrt{\frac{\beta\kappa^2}{2}}H_0\right)
\end{eqnarray}
\begin{itemize}\item
For $\omega<-1$, when $t\rightarrow -\infty$ one has $H\rightarrow
0$ (that is quantum effects are small at early times), on the other
hands, when   $t\rightarrow \bar{t}_s\equiv
t_s-\frac{2}{3(1+\omega)}
\sqrt{\frac{\beta\kappa^2}{2}}\left(\pi-\arctan\left(\sqrt{\frac{\beta\kappa^2}{2}}H_0\right)\right)$
one has $H\rightarrow \infty$, that is, the quantum effects don't
avoid the Big Rip singularity that appears at $\bar{t}_s$.

\item
For $\omega>-1$, when $t\rightarrow \infty$ one has $H\rightarrow 0$
(that is quantum effects are small at late times), on the other
hands, when  $t\rightarrow \bar{t}_s\equiv
t_s-\frac{2}{3(1+\omega)}\sqrt{\frac{\beta\kappa^2}{2}}
\left(\pi-\arctan\left(\sqrt{\frac{\beta\kappa^2}{2}}H_0\right)\right)$
one has $H\rightarrow \infty$, that is, the quantum effects don't
avoid the Big Bang singularity that appears at $\bar{t}_s$.
\end{itemize}

\item
Now we consider the case $\alpha=0$ and $\beta<0$. The solution of
equation (\ref{a13}) is given by
\begin{eqnarray}\label{a197}
-1/H(t)+\frac{1}{2H_+}\ln\left|\frac{H(t)-H_+}{H_0-H_+}\frac{H_0+H_+}{H(t)+H_+}\right|=
-\frac{3}{2}(1+\omega)(t-t_s),
\end{eqnarray}
with $H_+\equiv\sqrt{\frac{-2}{\beta\kappa^2}}$.

\begin{itemize}\item
In the case $\omega<-1$, for $H_0\in (0, H_+/\sqrt{2})$ when
$t\rightarrow -\infty$ on has $H\rightarrow 0$, and when
$H\rightarrow H_+/\sqrt{2}$ at $\bar{t}_s\equiv
t_s+\frac{2}{3(1+\omega)}\left(\sqrt{2}/H_+-\frac{1}{2H_+}\ln\left|\frac{
H_+/\sqrt{2}-H_+}{H_0-H_+}\frac{H_0+H_+}{
H_+/\sqrt{2}+H_+}\right|\right)$ one has
$\dot{H}(\bar{t}_s)=+\infty$, that is, these solutions are singular
(the scalar curvature, $R\equiv 6(2H^2+\dot{H})$, diverges). For
$H_0\in ( H_+/\sqrt{2}, H_+)$ when $t\rightarrow -\infty$ on has
$H\rightarrow H_+$, however at $\bar{t}_s$ $\dot{H}$ diverges. And
finally for $H_0\in ( H_+, \infty)$ there is a singularity at finite
time. Effectively , when $t\rightarrow -\infty$ on has $H\rightarrow
H_+$, and when $t\rightarrow t_s$ on has $H\rightarrow \infty$.
\item
In the opposite case $\omega>-1$, for $H_0\in (0, H_+/\sqrt{2})$
when $t\rightarrow \infty$ on has $H\rightarrow 0$, and when
$H\rightarrow H_+/\sqrt{2}$ at $\bar{t}_s\equiv
t_s+\frac{2}{3(1+\omega)}\left(\sqrt{2}/H_+-\frac{1}{2H_+}\ln\left|\frac{
H_+/\sqrt{2}-H_+}{H_0-H_+}\frac{H_0+H_+}{
H_+/\sqrt{2}+H_+}\right|\right)$ one has
$\dot{H}(\bar{t}_s)=+\infty$, that is, these solutions are singular.
For $H_0\in ( H_+/\sqrt{2}, H_+)$ when $t\rightarrow \infty$ on has
$H\rightarrow H_+$, however at $\bar{t}_s$ $\dot{H}$ diverges. And
finally for $H_0\in ( H_+, \infty)$ there is a singularity at finite
time. Effectively, when $t\rightarrow \infty$ on has $H\rightarrow
H_+$, and when $t\rightarrow t_s$ on has $H\rightarrow \infty$.
\end{itemize}
\end{enumerate}

\subsection{Empty universe}
Another simple case is the restriction to the invariant manifold
$\rho(t)\equiv 0$. In that case,  one only needs the  modified Friedmann
equation, that is, the following second order differential equation
\begin{eqnarray}\label{a198}
H^2=\kappa^2\alpha(3H^2\dot{H}+H\ddot{H}-\frac{1}{2}\dot{H}^2)-\frac{\kappa^2\beta}{2}
H^4
\end{eqnarray}

\begin{remark}
The solutions with $H>0$ and those with $H<0$ decouple. To see this,
we perform the change of variable $Z\equiv \dot{H}/H$ to make the
system no-singular at $H=0$, then at $H=0$ the system behaves
$\dot{Z}=Z^2/2$, this means that the solutions can't cross the axis
$H=0$.
\end{remark}

Equation (\ref{a198}) is an autonomous
second order differential equation, then since solutions are
invariant under time translations, the general solution is a
one-parameter family of solutions. Taking this into account,  we will prove that there is a one-parameter family of singular solutions.
First  we look for a particular singular solution, with the following behavior
$H(t)=\frac{C}{t-t_s}$ near the singularity. Inserting this
expression in (\ref{a198}), and retaining only the leading singular
terms one obtains
$C_{\pm}=\frac{3\alpha}{\beta}\left(-1\pm\sqrt{1+\frac{\beta}{3\alpha}}\right)$.
Here it is clear that we have to impose the condition
$\frac{\beta}{3\alpha}\geq-1$. In terms of the scale factor one
has $a(t)=a_0\left(\frac{t-t_s}{t_0-t_s}\right)^{C_{\pm}}$.
Then, for $\frac{\beta}{3\alpha}>0$, the solution with $C_+$ has a
singularity of the type $a(t_s)=0$, and the other one of the type
$a(t_s)=\infty$. However, when $-1\leq\frac{\beta}{3\alpha}<0$, for
both values of $C$, the solution satisfy $a(t_s)=0$.

Now we can prove that there is a one-parameter family of singular solutions
whose leading term is $H(t)=\frac{C_{\pm}}{t-t_s}$. To do this, we first transform the
differential equation (\ref{a198}) in a first order one performing the
change $u(H)=\dot{H}(t)$, then the equation becomes
\begin{eqnarray}\label{a199}
H^2=\kappa^2\alpha(3H^2u+Huu'-\frac{1}{2}\dot{H}^2)-\frac{\kappa^2\beta}{2}
H^4,
\end{eqnarray}
where $u'(H)\equiv du/dH$. Using the new variables
$H(t)=\frac{C_{\pm}}{t-t_s}$ has the form
$u=-\frac{1}{C_{\pm}}{H}^2$, then if the equation is linearized
about this point one obtains, for $H\rightarrow \pm\infty$, the
following general solution
\begin{eqnarray}\label{a200}
u_{linearized}=-\frac{1}{C_{\pm}}{H}^2+ K|H|^{1-3C_{\pm}}-\frac{C_{\pm}}{\kappa^2\alpha(3C_{\pm}-1)},\qquad \mbox{when}\qquad C_{\pm}\not= 1/3
\end{eqnarray} and
\begin{eqnarray}\label{a201}
u_{linearized}=-\frac{1}{C_{\pm}}{H}^2+ K-\frac{1}{3\kappa^2}\ln|H|,\qquad \mbox{when}\qquad C_{\pm}= 1/3,
\end{eqnarray}
where $K$ is an arbitrary constant.

It's clear that we have to impose $C_{\pm}>-1/3$. Then when
$-1<\frac{\beta}{3\alpha}<0$ or $\frac{\beta}{3\alpha}>15$, one has
$C_{\pm}>-1/3$, and then for both values one has a one-parameter
family of solutions. When $0<\frac{\beta}{3\alpha}<15$, for $C_+$, one
has a one-parameter family, however for $C_-$ one has to choose
$K=0$, that is, one only has a particular solution. \vspace{0.5cm}

Now we can perform the qualitative analysis. Firstly, note that when $\beta<0$ there
exist two de Sitter solutions
$H_{\pm}=\pm\sqrt{-\frac{2}{\kappa^2\beta}}$. Making the change of variable $p\equiv
\sqrt{|H|}$ (see \cite{aw86,wa85}), one obtains
\begin{eqnarray}\label{a202}
\frac{d}{dt}\left(\dot{p}^2/2+V(p)\right)=-3\epsilon p^2\dot{p}^2
\end{eqnarray}
where $\epsilon=sing(H)$ and
$V(p)=-\frac{p^2}{4\kappa^2\alpha}\left(1+\frac{\kappa^2\beta}{6}p^4\right)$. In
in the phase-space one has
\begin{eqnarray}\label{a203}\left\{\begin{array}{ccc}
\dot{p}&=& y\\
\dot{y}&=&-3\epsilon p^2y-V'(p).\end{array}\right.
\end{eqnarray}

We only consider the domain  $H>0$, because solutions with $H>0$ and $H<0$ decouple. The
point $p_+\equiv
\sqrt{H_+}$ is an extremum of the potential $V$, then, linearizing the
system (\ref{a21}), one obtains that for $\alpha<0$ the critical
point $(p_+,0)$ is a saddle point, and for $\alpha>0$ is a node
stable. From the form of the potential, and taking into account that
the system is dissipative in $H>0$, the
following results in the phase-space $(p,y)$, with $H>0$, are
obtained:

\begin{enumerate}\item
Case $\alpha>0$, $\beta>0$

We have $V<0$ in $(0,\infty)$ and $V(0)=0$. The $(0,0)$ is an
unstable critical point. The solutions are singular at early and
late times ($p\rightarrow \infty$). Only a solution is not singular
at late times, it is the trajectory that arrives at $p=0$ with zero
energy (it arrives at the point $(0,0)$), and only one if not
singular at early times, it starts from $p=0$ with zero energy (it
starts at $(0,0)$).
\item
Case $\alpha<0$, $\beta>0$

Now, $V>0$ in $(0,\infty)$ and $V(0)=0$. The $(0,0)$ is an stable
critical point, and solutions are only singular at early times. At
late times they approach to the stable critical point.

\item
Case $\alpha>0$, $\beta<0$

In this case, the system has two critical points. $(0,0)$ is an
unstable critical point, and $(p_+, 0)$ is  stable. Solutions are
only singular at early times. At late times they oscillate and
shrink around to the stable point, that is, $(p_+, 0)$ is a global
attractor. Moreover, there is a solution that ends at $(0,0)$, and
only a no singular solution that starts at $(0,0)$ (starts with zero
energy) and ends at $(p_+, 0)$.

\item
Case $\alpha<0$, $\beta<0$

This is the Starobinski model \cite{s80}. The system has two
critical points. $(0,0)$ is an stable critical point, and $(p_+, 0)$
is a saddle point. There are solutions that don't cross the axe
$p=p_+$, these solutions  are singular at early and late times, they
correspond to the trajectories that can't pass the top of the
potential. There are other solutions that cross twice the axe
$p=p_+$, they are also singular at early and late times, these
trajectories pass the top of the potential bounce at $p=0$ and pass
once again the top of the potential. There are solutions that cross
once the axe $p=p_+$, these solutions are singular at early times,
however at late times the solutions spiral and shrink to the origin,
these solutions pass the top of the potential once and then bounce
some times in $p=0$, shrinking to $p=0$. These last solutions has
the  asymptotic behavior described in the beginning of section 2.1.3.
Finally, there is only two unstable
no-singular solution, one goes from $(p_+, 0)$ to $(0,0)$, and the
other one is the de Sitter solution $(p_+, 0)$.

\end{enumerate}

\begin{remark}
Note that equation $(\ref{a16})$  remains the same with the change
$H(t)\rightarrow -H(-t)$, this means that solutions with $H<0$ are
the time reversal of the studied above.
\end{remark}

\begin{remark}
For $\frac{\beta}{3\alpha}<-1$ the values of $C_{\pm}$ are
complexes, this is due to the fact that the system cannot go to (or
come from) $p\rightarrow \infty$ monotonically, because the
dissipation effect is not large enough compared with the potential
force \cite{aw86}.
\end{remark}

\begin{remark}
The case $\omega=-1$ ($\rho=\mbox{constant}$ are invariant manifolds), is equivalent to the case of an empty universe with a cosmological constant. This case was studied with great detail in  \cite{aw86}.
\end{remark}

\subsection{The general case}
The best way to study the general case is to consider the system
\begin{eqnarray}\label{a204}\left\{\begin{array}{ccc}
\dot{H}&=&Y\\
\dot{Y}&=& \frac{1}{2\alpha H}\left({2H^2}/{\kappa^2}-\rho/3
-6\alpha H^2Y+\alpha Y^2+\beta H^4\right)\\
\dot{\rho}&=&-3H\rho(1+\omega).\end{array}\right.\end{eqnarray}

When $\beta<0$, the system has two critical points $(H_{\pm},0,0)$
whit $H_{\pm}=\pm\sqrt{-\frac{2}{\beta\kappa^2}}$.
The semi-plane $\rho=0, H>0$ (resp.  $\rho=0, H<0$) is an attractor
(resp. a "repeller") when $\omega>-1$, and the roles are interchanged
when $\omega<-1$.

What is important is to stress that in the case $\alpha<0$ there
isn't bouncing solutions, because at bouncing time, namely $t_b$,
one has $-\rho(t_b)/3+\alpha Y^2(t_b)=0$, what means $\rho(t_b)=0$,
but as we have seen in last section, $\rho=0$ is an invariant
manifold where trajectories with $H>0$ and those with $H<0$
decouple. For this reason in the case $\alpha<0$ there isn't stable
no-singular trajectories, the only  unstable no-singular solutions
are the ones that appear in the Starobinski model.

From this last paragraph one can concludes that, to obtain
no-singular solutions, the interesting case
 is $\alpha>0$. In fact, the interesting one is $\alpha>0$
and $\beta<0$.

In that case we can use the dimensionless variables $\bar{t}=H_+t$, $\bar{H}=H/H_+$
$\bar{Y}=Y/H_+^2$ and $\bar{\rho}=\frac{\kappa^2\rho}{6H_+^2}$, then the system becomes
\begin{eqnarray}\label{a205}\left\{\begin{array}{ccc}
{\bar{H}}'&=&\bar{Y}\\
{\bar{Y}}'&=& \frac{1}{2\alpha \bar{H}}\left(-\beta{\bar{H}^2}\beta\bar{\rho}
-6\alpha \bar{H}^2\bar{Y}+\alpha \bar{Y}^2+\beta \bar{H}^4\right)\\
{\bar{\rho}}'&=&-3\bar{H}\bar{\rho}(1+\omega),\end{array}\right.\end{eqnarray}
where $'$ denotes the derivative with respect the time $\bar{t}$.

In these variables the critical points are $(\pm 1,0,0)$. The
linearized system at $(1,0,0)$ has eigenvalues
$\lambda_{\pm}=-3/2\left(1\pm\sqrt{1+\frac{4\beta}{3\alpha}}\right)$
and $\lambda_3=-3(1+\omega)$. Since $\Re{\lambda}_{\pm}<0$ and
$\bar{\rho}\equiv 0$ is an invariant manifold, all solutions in that
semi-plane with $\bar{H}>0$ go asymptotically towards  this critical
point. The eigenvector
$\vec{v}_3=(1,-3(1+\omega),18\omega(1+\omega)\alpha/\beta-2)$
corresponds to the eigenvalue   $\lambda_3$, then for $\omega<-1$,
there is a solution that escapes to the de Sitter expanding phase
following the direction of the vector $\vec{v}_3$. On the other
hands, when $\omega>-1$ the critical point is an attractor. At the
other critical point the eigenvalues are
$\lambda_{\pm}=-3/2\left(1\pm\sqrt{-1+\frac{4\beta}{3\alpha}}\right)$
and $\lambda_3=3(1+\omega)$. Since $\Re{\lambda}_{\pm}>0$ and
$\bar{\rho}\equiv 0$ is an invariant manifold, all solutions in that
semi-plane with $\bar{H}<0$ escape from this critical point. The eigenvector
$\vec{v}_3=(1,3(1+\omega),-18\omega(1+\omega)\alpha/\beta+2)$
corresponds to the eigenvalue   $\lambda_3$, then for $\omega<-1$,
there is a particular solution that goes asymptotically towards  the de
Sitter contracting phase following the direction of the vector
$\vec{v}_3$. On the other hands, when $\omega>-1$ the critical point
is a repeller.

\vspace{0.5cm}

Now we look for singular solutions of the form
$\bar{H}=C/(\bar{t}-\bar{t}_s)$, $\bar{t}\rightarrow
\bar{t}_s^{\pm}$. Inserting this value of the Hubble parameter in
the conservation equation one obtains
$\bar{\rho}(t)=\bar{\rho}_0|\bar{t}-\bar{t}_s|^{-3C(1+\omega)}$,
where $\bar{\rho}_0$ has to be a positive parameter. Now , when $\omega<-1$,
inserting the Hubble parameter and the energy density in the
modified Friedmann equation one gets, once again, the values of $C$
obtained in section 4.3, that is,
$C_{\pm}=\frac{3\alpha}{\beta}\left(-1\pm\sqrt{1+\frac{\beta}{3\alpha}}\right)$,
because in that case the energy goes to zero when
$\bar{t}\rightarrow \bar{t}_s^{\pm}$, and then, one obtains the same
kind of result that in the case $\bar{\rho}\equiv 0$, the only difference is that
now one has a two-parameter family of singular solutions ($\bar{\rho}_0$ is a free parameter, and the
general solution of the system (\ref{a23}) is a two-parameter family due to the time invariance under translations).

The case $\omega>-1$ is more involved. We summarize the results:
\begin{enumerate}
\item When $-1<\frac{\beta}{3\alpha}<0$
\begin{itemize}
\item  If $C_-<\frac{4}{3(1+\omega)}$,  there are two
two-parameter families with $\bar{H}=C_{\pm}/(\bar{t}-\bar{t}_s)$
and $\bar{\rho}_0$ free parameter.
\item If $C_-=\frac{4}{3(1+\omega)}$, there is a two-parameter
family with $\bar{H}=C_{+}/(\bar{t}-\bar{t}_s)$ and $\bar{\rho}_0$
free parameter, and a one-parameter family with
$\bar{H}=C_{-}/(\bar{t}-\bar{t}_s)$ and $\bar{\rho}_0\equiv 0$.
\item If $C_+<\frac{4}{3(1+\omega)}<C_-$, there is a two-parameter
family with $\bar{H}=C_{+}/(\bar{t}-\bar{t}_s)$ and $\bar{\rho}_0$
free parameter, a one-parameter family with
$\bar{H}=C_{-}/(\bar{t}-\bar{t}_s)$ and $\bar{\rho}_0\equiv 0$, and
a one-parameter family with
$\bar{H}=\frac{4}{3(1+\omega)(\bar{t}-\bar{t}_s)}$ and
$\bar{\rho}_0=-\frac{16}{9(1+\omega)^2}\left(\frac{16}{9(1+\omega)^2}+
\frac{6\alpha}{\beta}\frac{4}{3(1+\omega)}
-\frac{3\alpha}{\beta}\right)$.
\item If $\frac{4}{3(1+\omega)}\leq C_+ $, there are two
one-parameter families with $\bar{H}=C_{\pm}/(\bar{t}-\bar{t}_s)$
and $\bar{\rho}_0=0$.
\end{itemize}

\item When $-1=\frac{\beta}{3\alpha}$
\begin{itemize}
\item  If $1<\frac{4}{3(1+\omega)}$, there is a
two-parameter family with $\bar{H}=1/(\bar{t}-\bar{t}_s)$ and
$\bar{\rho}_0$ free parameter.
\item  If $1\geq \frac{4}{3(1+\omega)}$, there is a
one-parameter family with $\bar{H}=1/(\bar{t}-\bar{t}_s)$ and
$\bar{\rho}_0=0$.
\end{itemize}
\item When $-1>\frac{\beta}{3\alpha}$, there aren't singular
solutions of the form $\bar{H}=C/(\bar{t}-\bar{t}_s)$.
\end{enumerate}

\begin{remark}
This result can be obtained in an equivalent way inserting in
equation (\ref{a194}) the function $H=C/(t-t_s)$. Then, retaining the
leading singular term, one obtains the values $C_{\pm}$ and
$\frac{4}{3(1+\omega)}$. Finally, transforming the differential
equation in  a second order one in the same way as we have done in
Section IV,  and linearizing around  the singular behaviors obtained
above, one can see the form and the number of parameters that depend
the different families of singular solutions.
\end{remark}

To understand this summary we perform the change of variable
$p\equiv\sqrt{|H|}$. Then the modified Friedmann equation becomes
\cite{wa85}
\begin{eqnarray}\label{a206}
\frac{d}{dt}\left(\dot{p}^2/2+\tilde{V}(p)\right)=-3\epsilon
p^2\dot{p}^2+\frac{3\epsilon}{24\alpha}(1+\omega)\rho,
\end{eqnarray}
where
$\tilde{V}(p)=-\frac{p^2}{4\kappa^2\alpha}\left(1+\frac{\kappa^2\beta}{6}p^4\right)-\frac{\rho}{24\alpha
p^2}$, and $\epsilon\equiv sgn(H)$.

The case $\omega<-1$ is clear. Since $\bar{\rho}\rightarrow 0$ at
$\bar{t}=\bar{t}_s$, one essentially obtains the same results as
Section IV. However, for $\omega>-1$, on the right hand side of
equation (\ref{a206}) one term is  dissipative  and the other one is
anti-dissipative, moreover, in this case both terms diverge at
$\bar{t}=\bar{t}_s$. Then if one looks for singular solutions of the
form $\bar{H}=C/(\bar{t}-\bar{t}_s)$, the first term in the right
hand side of (\ref{a24}) has to be dominant. And since this term is
of the order $1/(\bar{t}-\bar{t}_s)^4$, and the other one is of the
order $1/|\bar{t}-\bar{t}_s|^{3C(1+\omega)}$, they will appear all
the situations described above.

It is also interesting to understand the form of the potential
$\tilde{V}$ (its picture appears in figure 3 of ref. \cite{wa85}). It only
has a zero at the point
$p_0=\left(3/2\right)^{1/4}\left(1+\sqrt{1+\frac{4}{3}\bar{\rho}}\right)^{1/4}$,
and two critical points at
$p_{\pm}=\left(\frac{1\pm\sqrt{1-{4}\bar{\rho}}}{2}\right)^{1/4}$
($p_-<p_+$). Then for $\bar{\rho}>1/4$ there aren't critical points,
and the potential is strictly increasing from $-\infty$ to $\infty$.
For $\bar{\rho}<1/4$, the potential satisfy $\tilde{V}(0)=-\infty$ ,
$\tilde{V}(\infty)=\infty$ and has a relative maximum at $p_-$ and a
relative minimum at $p_+$ (a hollow). For very small values of
$\bar{\rho}$ at $p_-$ one has $\bar{H}^2\cong\bar{\rho}$, that is,
the system is nearly to the Friedmann phase, and at $p_+$ one has
$|\bar{H}|\cong 1$, that is, the system is near to the de Sitter
phase.

Next step is to find solutions  that approximate to the Friedmann one when $|t|\rightarrow\infty$
(see \cite{fhh79} for the radiation case, i.e., $\omega=1/3$). To do that, we
consider equation (\ref{a194}) in the dimensionless variables introduced above, and we reduce the order performing
the change of variable $u(y)=\bar{H}(\bar{t})$ where $y=\bar{H}$. The obtained equation is:
\begin{eqnarray}\label{a207}
2\beta u+3(1+\omega)\beta(y^2-y^4)+(18\alpha(1+\omega)-4\beta)y^2 u+6\alpha(2+\omega)y\dot{u}u
+3\alpha(3+\omega)u^2+2\alpha(\ddot{u}u^2+\dot{u}^2u)=0,
\end{eqnarray}
where now $\dot{u}\equiv du/dy$. Since the Friedmann solution in
these variables is $u_{F}= -\frac{3}{2}(1+\omega)y^2$, the
linearized equation about this point ($u_{linearized}=u_F+h$) is
\begin{eqnarray}\label{a208}
 \ddot{h}+{2}{y}^{-1}\frac{\omega}{1+\omega}\dot{h}+\left(\frac{4\beta}{9\alpha(1+\omega)^2}y^{-4}+
Ay^{-2}\right)h+B=0,
\end{eqnarray}
 where $A$ and $B$ are some constants depending on the parameters $\alpha$, $\beta$ and $\omega$.

The idea to solve this equation is to take into account that for large values of $|t|$ (small values of $y$), one can
disregard the term  $Ay^{-2}$. The homogeneous equation is solved performing the change
$h=|y|^{-\frac{\omega}{1+\omega}}z$, then one obtains:
\begin{eqnarray}\label{a209}
 \ddot{z}+\frac{4\beta}{9\alpha(1+\omega)^2}y^{-4}z=0,
\end{eqnarray}
 that we solve using the WKB approximation (see \cite{a83} page $276$). Consequently the homogeneous equation has the two independent solutions
\begin{eqnarray}\label{a210}
 h_{homogeneous,\pm}(y)=y^{1/(1+\omega)}exp\left({\pm\frac{2}{3(1+\omega)}\sqrt{-\beta/\alpha}\frac{1}{y}}\right).
\end{eqnarray}
A particular solution is obtained using power series. It leading term is
\begin{eqnarray}\label{a211}
 h_{particular}(y)=\frac{9\alpha(1+\omega)^2}{4\beta}By^4.
\end{eqnarray}

Then the general solution of the linearized equation is approximately
\begin{eqnarray}\label{a212}
 u_{linearized}=u_F+Kh_{homogeneous,\pm}(y)+h_{particular}(y) \quad \mbox{for}\quad \mp y>0,
\end{eqnarray}
where $K$ is an arbitrary parameter, that is, we have proved that
there is  a one-parameter family of solutions that approximate to
the Friedmann one for large values of $|t|$.

Once we have seen these preliminary results, we can describe
qualitatively the behavior of the no-singular solutions. We start
with the case $w<-1$. We have seen that there is a one-parameter
family of solutions that at early times are in the expanding
Friedmann phase, and we have to look for  no-singular
solutions  that match, at late time, with that family. The only
no-singular solutions at early times are: a one-parameter family
that approaches asymptotically to the contracting Friedmann phase,
and a particular solution that goes asymptotically towards the
contracting de Sitter phase, following the direction
$(1,3(1+\omega),-18\omega(1+\omega)\alpha/\beta+2)$. From the system
(\ref{a24}) it is easier to understand the dynamics. First, at early
times the system is at the point $p_-$ with $\bar{H}>0$. Then it
leaves this expanding Friedmann state and rolls down either to the
right or to the left. In the former case, the universe approaches to
an expanding de Sitter phase (the relative minimum $p_+$). However
since $\bar{\rho}$ is an increasing function with time, the critical
points  will disappear and the potential will be an increasing
function with $p$, this means that the universe rolls down to $p=0$,
that is, it bounces and enters in a decreasing phase $\bar{H}<0$.
Then it can arrive asymptotically at the points $p_-$ or $p_+$, (the
no-singular solutions at late time), or it bounces many times in
order to have enough energy in $\bar{H}<0$, to arrive at $p=\infty$
(singular solution). This last behavior can be easily understood, if
 one takes  first into account that for $\bar{H}>0$ (resp.
$\bar{H}<0$) the system is dissipative (resp. anti-dissipative), and
second that the energy of the system changes its sing when it
bounces (see equation $(12)$ of \cite{aw86}).

Finally,  from the behavior of the no-singular solutions at late
time, one can deduce that one has to very fine tune the initial conditions and
the parameters
$\alpha$ and $\beta$ in order to obtain no-singular solutions that
match these late time no-singular behaviors with the expanding
Friedmann stage at early times, because these families of solutions aren't
general solutions (a two-parameter family).

On the other hand when $\omega>-1$, we also have a one-parameter
family of solutions that at late times are in the expanding
Friedmann phase (in terms of the variable $p$ this corresponds to
the point $p_-$ and $H>0$), and we have to look for  no-singular
solutions  that match, at early time, with that family. The only
no-singular solutions at early times are: a one-parameter family
that leaves the contracting Friedmann phase,  in terms of the
variable $p$, this means, that the system leave the  relative
maximum $p_-$ with $\bar{H}<0$, and rolls down to the right or to the
left, but in all cases, since the energy density is an increasing
function of time in this region, the system goes to $p=0$, i.e., it
bounce and starts an expanding phase. The other no-singular solution
at early times is a two-parameter solution (a general solution) that
leaves the contracting de Sitter phase, in terms of the variable $p$,
this means, that the system stars at $p_+$ with $\bar{H}<0$, and
then due to the anti-dissipation (at early times in $\bar{H}<0$ the
energy density is very small and the dominant term is the first one
on the right hand side of  equation (\ref{a206})) the system is
released from the hollow and rolls down towards the region $\bar{H}>0$.

From these no-singular early time behavior one can conclude that, in
order to match these early time no-singular solutions with the
expanding Friedmann phase at late time, one has to fine tune the
initial condition. And depending on the values of the parameters
$\alpha$ and $\beta$ we will obtain different kinds of connections.
For example, in \cite{a83}, different numerical calculations have been
done in the radiation case, and they show the different connections
in terms of both parameters.

\vspace{1cm}

\noindent{\bf Acknowledgments.} This investigation has been
supported in part by MICINN (Spain), project MTM2008-06349-C03-01
and by AGAUR (Generalitat de Ca\-ta\-lu\-nya), contract
2005SGR-00790.

\end{document}